\documentclass[prc,twocolumn,a4paper,floatfix,nofootinbib,preprintnumbers,superscriptaddress]{revtex4-1}

\usepackage[utf8]{inputenc}
\usepackage{graphicx,psfrag}
\usepackage{mathrsfs}
\usepackage{amsmath,amsfonts,amssymb}
\usepackage{multirow} 
\usepackage{diagbox}
\usepackage{comment}
\usepackage{xcolor}
\usepackage{enumerate}
\usepackage{booktabs}
\usepackage{lineno}\usepackage{lipsum}
\usepackage{mathtools}
\usepackage{color,soul}

\usepackage{hyperref}
\hypersetup{
    colorlinks = true,
    linkcolor = {blue},
    citecolor = {blue},
    urlcolor = {blue},
    linkbordercolor = {white},
    }

\usepackage{color}
\definecolor{cyan}{rgb}{0,0.9,0.9}
\definecolor{orange}{rgb}{0.9,0.5,0}
\definecolor{magenta}{rgb}{1,0,1}
\definecolor{purple}{rgb}{0.8,0.4,0.8}
\definecolor{gray}{rgb}{0.8242,0.8242,0.8242}
\definecolor{mgreen}{rgb}{0.1,0.8,0.1}

\usepackage[normalem]{ulem}

\newcommand{\nh}{n_0}
\begin{document}
\preprint{LA-UR-23-29118} 
\preprint{NP3M-P2300017}

\title{Probing Quarkyonic Matter in Neutron Stars with the Bayesian Nuclear-Physics Multi-Messenger Astrophysics Framework}

\author{Peter~T.~H. \surname{Pang}}
\email{t.h.pang@uu.nl}
\affiliation{Nikhef, Science Park 105, 1098 XG Amsterdam, The Netherlands}
\affiliation{Institute for Gravitational and Subatomic Physics (GRASP), Utrecht University, Princetonplein 1, 3584 CC Utrecht, The Netherlands}

\author{Lars \surname{Sivertsen}}
\email{lars@iastate.edu}
\affiliation{Department of Physics and Astronomy, Iowa State University, Ames, IA 50010}

\author{Rahul \surname{Somasundaram}}
\affiliation{Theoretical Division, Los Alamos National Laboratory, Los Alamos, NM 87545, USA}
\affiliation{Department of Physics, Syracuse University, Syracuse, NY 13244, USA}

\author{Tim \surname{Dietrich}}
\affiliation{Institute for Physics and Astronomy, University of Potsdam, D-14476 Potsdam, Germany}
\affiliation{Max Planck Institute for Gravitational Physics (Albert Einstein Institute), Am M\"uhlenberg 1, Potsdam 14476, Germany}

\author{Srimoyee \surname{Sen}}
\affiliation{Department of Physics and Astronomy, Iowa State University, Ames, IA 50010}

\author{Ingo \surname{Tews}}
\affiliation{Theoretical Division, Los Alamos National Laboratory, Los Alamos, NM 87545, USA}

\author{Michael Coughlin}
\affiliation{School of Physics and Astronomy, University of Minnesota, Minneapolis, Minnesota 55455, USA}

\author{Chris \surname{Van Den Broeck}}
\affiliation{Nikhef, Science Park 105, 1098 XG Amsterdam, The Netherlands}
\affiliation{Institute for Gravitational and Subatomic Physics (GRASP), Utrecht University, Princetonplein 1, 3584 CC Utrecht, The Netherlands}

\date{\today}

\begin{abstract} 
The interior of neutron stars contains matter at the highest densities realized in our Universe. 
Interestingly, theoretical studies of dense matter, in combination with the existence of two solar mass neutron stars, indicate that the speed of sound $c_s$ has to increase to values well above the conformal limit ($c_s^2 = 1/3$) before decreasing again at higher densities. 
The decrease could be explained by either a strong first-order phase transition or a cross-over transition from hadronic to quark matter. 
The latter scenario leads to a pronounced peak in the speed of sound reaching values above the conformal limit, naturally explaining the inferred behavior.
In this work, we use the Nuclear-Physics Multi-Messenger Astrophysics framework \textsc{NMMA} to compare predictions of the quarkyonic matter model with astrophysical observations of neutron stars, with the goal of constraining model parameters.
Assuming quarkyonic matter to be realized within neutron stars, we find that there can be a significant amount of quarks inside the core of neutron stars with masses in the two solar mass range, amounting to up to $\approx 0.13M_\odot$, contributing $\approx 5.9\%$ of the total mass.
Furthermore, for the quarkyonic matter model investigated here, the radius of a $1.4M_\odot$ neutron star would be $13.44^{+1.69}_{-1.54} (13.54^{+1.02}_{-1.04})$ km, at $95\%$ credibility, without (with) the inclusion of AT2017gfo.
\end{abstract}

\maketitle

\section{Introduction}
\label{sec:intro}

Measurements of neutron stars' masses, radii, and deformabilities serve as probes of the nuclear equation of state (EOS) of neutron-rich matter at high density and low temperature ~\cite{ozel:2016oaf}. 
This part of the QCD phase diagram can not be accessed in lattice QCD simulations due to the sign problem, which plagues the importance sampling of gauge configurations. 
Thus, neutron star measurements remain one of the few tools to explore the behavior of dense cold matter. For instance, the feasibility of constraining the EOS via measurement of neutron stars' tidal deformabilities with gravitational waves has been demonstrated~\cite{Flanagan:2007ix,DelPozzo:2013ala}.
Constraints on the radius $R$ of neutron stars obtained from gravitational-wave data suggest that $R<13.5$ km~\cite{Annala:2017llu, Abbott:2018exr, De:2018uhw,Tews:2018iwm}, which in turn implies that the EOS of matter is soft at nuclear densities, of the order of the nuclear saturation density $n_{\rm sat}\simeq 0.16$~fm$^{-3}$. 
On the other hand, observations of heavy pulsars indicate that the EOS needs to be sufficiently stiff at higher densities 
in order to support neutron stars with masses of more than twice the solar mass~\cite{Watts:2016uzu,Demorest:2010bx,Antoniadis:2013pzd,PhysRevC.103.045808,Raaijmakers_2020}.  

This soft-stiff behavior of the EOS also reveals an interesting feature of the speed of sound ($c_s$) in dense matter as a function of density. 
The speed of sound can be calculated with uncertainty estimates at nuclear densities using chiral effective theory ($\chi$EFT)~\cite{Hebeler:2013nza,Tews:2018kmu,Drischler:2020hwi,Keller:2022crb} and at very high density using perturbative QCD~\cite{Gorda:2018gpy,Gorda:2021znl,Gorda:2023mkk}. 
Thus, we know that the speed of sound $c_s$ is small, $c_s^2\ll 1$, at low densities, and approaches the conformal bound of $c_s=1/3$ from below at very high densities. 
At intermediate densities, where neither chiral EFT nor perturbative QCD are applicable, the speed of sound could, in principle, be a monotonic or nonmonotonic function of baryon density. 
The soft-stiff EOS obtained from neutron-star measurements reveals that the speed of sound, in fact, is likely a non-monotonic function of baryon density, violating the conformal bound at a few times the nuclear saturation density~\cite{Bedaque:2014sqa,Tews:2018kmu}. 
More specifically, $c_s^2$ increases monotonically from nuclear saturation density to a few times the nuclear saturation density, overshooting the conformal bound of $1/3$ and reaching a peak at intermediate densities.
With even larger densities, $c_s^2$ then decreases below the conformal bound, reaching a minimum and eventually increasing to asymptotically approach the conformal bound from below.  
Given that we do not have the tools to obtain the EOS of dense matter from direct calculations in QCD, it is important to assess various models which can replicate this non-monotonic behavior of the speed of sound and simultaneously describe mass and radius measurements of neutron stars. 

One such model is ``quarkyonic matter''~\cite{McLerran:2018hbz,Jeong:2019lhv,Sen:2020qcd} in which a special arrangement of nucleons and quarks in a combined Fermi sphere forces a peak in the sound speed. 
In the quarkyonic model, the nucleon and quark degrees of freedom (DOF) are described by a single Fermi distribution function, as illustrated in Fig.~\ref{fermi_sphere}. 
At high densities, the Fermi momentum for the baryons is large and the DOF deep within the Fermi sphere are Pauli blocked. 
Creating a particle-hole excitation from deep within the Fermi sea requires large energy and momentum, and so these DOF can be regarded as weakly interacting. 
Since QCD is asymptotically free, we expect the existence of quark matter at high densities where quarks behave as nearly free particles, thus motivating treating the DOF deep within the Fermi Sphere as quarks. 
The DOF near the Fermi surface can, however, be excited with low energy and momentum transfers. 
Thus, confining forces should remain important. 
This motivates treating the DOF near the Fermi surface as nucleons arising from quark correlations.

At low densities, the radius of the inner quark Fermi sphere in the quarkyonic model is zero. 
As the density increases, the inner quark Fermi sphere starts forming at some threshold density. 
The peak in the speed of sound arises right at those densities where the inner quark Fermi sphere first appears.

Interestingly, the introduction of quarks in other models is typically accompanied by a first-order phase transition. A first-order transition forces the pressure gradient to drop to zero, leading to a lower pressure than one would encounter in the absence of a phase transition. In the quarkyonic case, however, the transition to quark matter happens through a crossover. Not only does this mean that the pressure gradient is non-zero during  the transition, it also does not decrease nearly as fast after the peak in the speed of sound as it does in the first order transition case. This keeps the EOS stiff for a larger range of densities. The quarkyonic model, therefore, differs from other models of quark matter in this regard.

Although quarkyonic models have been successful in producing a peak in the speed of sound, there have not been any studies to infer the quarkyonic-matter model parameters from astrophysical data. 
In particular, given an unconstrained quarkyonic model, we would like to investigate what we can learn about the model parameters from astrophysical data and what the quark content is in neutron stars.  
In this paper, we answer these questions using Bayesian analysis and adopting a quarkyonic excluded-volume model proposed in Ref.~\cite{Sen:2020qcd}.

The astrophysical observations considered here are the gravitational-wave detections of the binary neutron star mergers GW170817~\cite{TheLIGOScientific:2017qsa} and GW190425~\cite{Abbott:2020uma} by Advanced LIGO~\cite{LIGOScientific:2014pky} and Advanced Virgo~\cite{VIRGO:2014yos}, the GW170817-associated electromagnetic counterparts AT2017gfo~\cite{GBM:2017lvd}, the NICER observations on PSR J0030+0451\cite{Miller:2019cac,Riley:2019yda} and PSR J0740+6620~\cite{Miller:2021qha,Riley:2021pdl}, and the radio observations on PSR J0348+4032~\cite{Antoniadis:2013pzd} and PSR J1614-2230~\cite{Arzoumanian:2017puf}.

This paper is organized as follows: We start by describing the employed quarkyonic-matter model~\cite{Sen:2020qcd} in Section~\ref{quarkyonic EOS}. 
We then describe our Bayesian inference approach in Section~\ref{bayesian inference}, and give a short description of our implementation in Section~\ref{implementation}. 
Finally, we present our results followed by our conclusions in Sections~\ref{sec:results} and~\ref{sec:conclusion}.

\section{Methods}
\label{sec:methods}

\subsection{Quarkyonic-Matter EOS}\label{quarkyonic EOS}

A non-dynamic quarkyonic matter model was first proposed in Ref.~\cite{McLerran:2018hbz}, followed by a dynamic model for quarkyonic matter introduced in Ref.~\cite{Jeong:2019lhv}. 
In this dynamic model, both hadrons and quarks appear as quasi-particles and are described using a single Fermi distribution which has an outer shell of nucleons and an inner sphere of quarks. 
The baryon density of nucleons $n_{\rm B}^{\rm N}$ for a particular configuration of quarkyonic matter is computed by evaluating the momentum space volume of the outer Fermi shell, whereas the quark baryon density $n_{\rm B}^{\rm Q}$ is computed by evaluating the volume of the inner sphere. 
The total baryon density is given by the sum of the two: 
$n_{\rm B} = n_{\rm B}^{\rm N}+n_{\rm B}^{\rm Q}$. 
Thus, for a fixed baryon density, one can vary the shell width made of the nucleons (or the radius of the inner quark Fermi sphere) while maintaining a constant $n_{\rm B}$, to probe different configurations of quarkyonic matter for that particular total baryon density. 
The dynamical model in Ref.~\cite{Jeong:2019lhv} also introduces an energy density functional for the systems, defined as the sum of individual energy density functionals of the nucleons and quarks. 
By minimizing the energy density functional for the quarkyonic Fermi distribution with respect to the shell radius of the quarkyonic phase space, one can find the equilibrium quarkyonic matter configuration for a fixed baryon density. 
One can then construct the EOS of dense matter by computing the energy density for the equilibrium configuration as a function of density. 
Note that the dynamical model of Ref.~\cite{Jeong:2019lhv} considers only symmetric nuclear matter. 
This model was extended to account for the astrophysically more relevant system of pure neutron matter in Ref.~\cite{Sen:2020qcd}.
We use this extended model in this work.
In the next subsection, we describe the construction of the relevant energy density functionals, starting with the neutrons, and then moving on to the quarks.

\begin{figure}
    \centering
    \includegraphics[width=\linewidth]{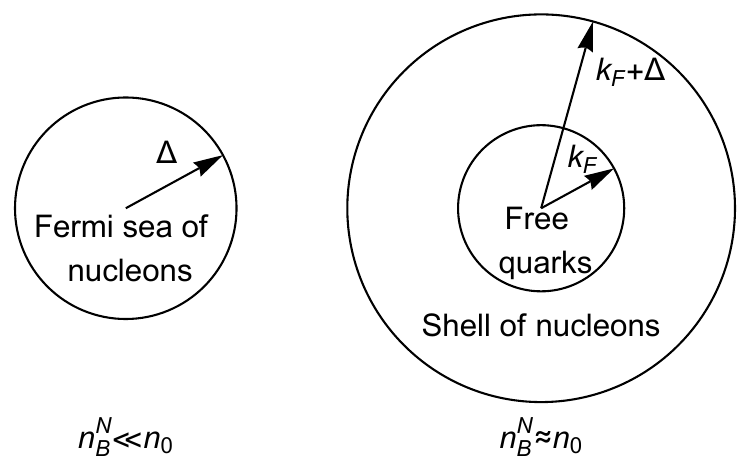}
    \caption{Illustration of the Fermi sphere for quarkyonic matter at low baryon number densities before the quark onset (left panel) and at high densities after the quark onset (right panel).
    The quark Fermi momentum in dynamic quarkyonic-matter models well below $n_0$ is negligible and not shown in this figure.}
    \label{fermi_sphere}
\end{figure}

\subsubsection{Neutron part of the EOS}

In the excluded volume model for quarkyonic neutron matter, we consider neutrons to be hard spheres of volume $v_h$ exhibiting two spin DOFs. This allows us to define an effective density for neutrons. 
If we insert $N$ neutrons in a volume $V$, then the unoccupied volume $V_\textrm{ex}$ is given by
\begin{equation}
    V_\textrm{ex} = V-Nv_h\equiv V-\frac{N}{\nh}\,.
\end{equation}
In the above, we have defined the hard core density of neutrons as $n_0\equiv v_h^{-1}$. If we now add another neutron to this $N$-neutron system, this extra neutron can only occupy a volume $v_h$ within the available volume $V_\textrm{ex}$. 
This gives rise to an increased effective density for the neutrons, which we will denote as the neutron excluded volume density $n^\textrm{N}_\textrm{ex}$:
\begin{equation}
    n^\textrm{N}_\textrm{ex} \equiv \frac{N}{V_\textrm{ex}} = \frac{V n_\textrm{B}^\textrm{N}}{V-\frac{N}{\nh}}= \frac{n_\textrm{B}^\textrm{N}}{1-\frac{N}{V \nh}} = \frac{n_\textrm{B}^\textrm{N}}{1-\frac{n_\textrm{B}^\textrm{N}}{\nh}}\,.
\end{equation}
This implies that the neutron density $n_{\text{B}}^{\text{N}}$ is constrained to be lower than the hard-core density $\nh$ since otherwise, the volume occupied by neutrons would be larger than $V$. 
Because of this, the energy density associated with nucleons will increase faster than that of a free Fermi gas as a function of baryon number density. 
If we then consider quarks to be free point particles, as we will describe in the next subsection, the sum of the energy density contributions from quarks and neutrons will eventually be lower if some of the baryon density is stored in quark DOF compared to a pure neutron phase. 
This is the main idea behind this excluded volume model: as we increase the number of neutrons in a constrained volume, it eventually becomes energetically favorable for some of those neutrons to split into their constituent quarks, leading to a smooth crossover between a hadronic phase and a quark matter phase. 

Keeping in mind that the neutron excluded volume density is the effective density for the neutrons, we can write $n_\textrm{ex}^\textrm{N}$ in terms of neutron Fermi momentum $\Delta$,
\begin{equation}
    n_\textrm{ex}^\textrm{N} = 2\int_0^{\Delta} \frac{\textrm{d}^3k}{(2\pi)^3}, 
\end{equation}
where the factor of two accounts for the neutron's two spin DOF. 
From this, it follows that the non-interacting pure neutron energy density functional for hard-core neutrons is given by
\begin{equation}
    \epsilon_\textrm{N} = 2\Big(1-\frac{n_\text{B}^\textrm{N}}{\nh}\Big)\int_0^{\Delta}\frac{\textrm{d}^3 k}{(2\pi)^3}\sqrt{M_\textrm{N}^2+k^2},
    \label{en}
\end{equation}
where $M_\textrm{N}$ is the neutron mass. 

To properly describe neutron stars, we will have to include neutron interactions in our energy density functional.
For this, we use a potential ansatz inspired by Ref.~\cite{McLerran:2018hbz}, and include a nuclear potential of the form 
\begin{equation}
    V(n_\textrm{B}^\textrm{N}) = \tilde{a} n_\textrm{B}^\textrm{N} \Big(\frac{n_\textrm{B}^\textrm{N}}{\rho_0}\Big)+\tilde{b} n_\textrm{B}^\textrm{N} \Big(\frac{n_\textrm{B}^\textrm{N}}{\rho_0}\Big)^2\,,
\end{equation}
where $\rho_0$ is the saturation density, and $\tilde{a}$ and $\tilde{b}$ are free parameters.
The parameters $\tilde{a}<0$ and $\tilde{b}>0$ are to mimic the short-range repulsion and long-range attraction between neutrons, respectively, at densities $n_\textrm{B}\lesssim 2\rho_0$. 

However, this is not sufficient to reproduce the EOS of low-density neutron matter. 
The excluded volume energy density for the neutrons in Eq.~\eqref{en} is desirable at high densities to force quark onset, but the excluded volume model has the undesirable feature of significantly modifying the low-density neutron EOS as well. 
In order to correct for this effect, Ref.~\cite{Sen:2020qcd} proposed the following approach: We introduce a modified potential $\tilde{V}(n_{\rm B}^{\rm N})$, 
and the energy density functional 
\begin{equation}
    \epsilon_\textrm{N} = 2\Big[1-\Big(\frac{n_\textrm{B}^\textrm{N}}{\nh}\Big)\Big]\int_0^\Delta\frac{\textrm{d}^3k}{(2\pi)^3}\sqrt{M_\textrm{N}^2+k^2}+\tilde{V}(n_\textrm{B}^\textrm{N}),\label{e1}
\end{equation}
to minimize the effect of the excluded volume at low density. $\tilde{V}$ is designed such that in a low-density expansion in powers of ${n_{\rm B}^{\rm N}}/{\nh}$, the excluded volume energy density of Eq. \ref{e1} mimics the energy density of low density neutron matter given by 
\begin{equation}
\epsilon=2\int_0^{k_F} \frac{d^3k}{(2\pi)^3}\sqrt{k^2+M_N^2}+V(n_N^N)
\end{equation}
up to corrections that go as some high power in the expansion parameter.
Note that, $n_{\rm B}^N=2\int_0^{k_F}\frac{d^3k}{(2\pi)^3}$. 
The model proposed in Ref.~\cite{Sen:2020qcd} demands that the two energy densities match upto corrections that go as the $(n_{\rm B}^N/n_0)^3$.
The corresponding expression for $\tilde{V}$ is given by
\begin{align}
    \tilde{V}(n_\textrm{B}^\textrm{N}) =V(n_\textrm{B}^\textrm{N}) &- \frac{2}{3}\frac{(3\pi^2 n_\textrm{B}^\textrm{N})^{5/3}}{10\pi^2 M_\textrm{N}}\Big(\frac{n_\textrm{B}^\textrm{N}}{\nh}\Big)
    \nonumber
    \\
    &- \frac{5}{9}\frac{(3\pi^2 n_\textrm{B}^\textrm{N})^{5/3}}{10\pi^2 M_\textrm{N}}\Big(\frac{n_\textrm{B}^\textrm{N}}{\nh}\Big)^2\,.
\end{align}

\subsubsection{Free quark part of the EOS}\label{free_quarks}

We will now discuss the energy density functional for the quarks. 
We treat the quarks as almost free point particles of mass $M_{\rm q}=M_{\rm N}/N_c=M_{\rm N}/3$, where $N_c$ is the number of quark colors and fixed at $N_c=3$. 
The only effect of interactions of quarks is taken into account through the constituent quark mass, $M_{\rm q}$. 
Here, we include two species of quarks, up and down, with two spin DOFs. 
We can then express the contribution to the baryon density from quarks in terms of the quark Fermi momentum $k_\textrm{F}^\textrm{q}$, so that
\begin{equation}
    n_\textrm{B}^\textrm{q} = 2\int_0^{k_\textrm{F}^\textrm{q}}\frac{\textrm{d}^3 k}{(2\pi)^3},
\end{equation}
where q = \{u,d\} denotes the quark flavor, and the factor of two again comes from the spin DOF. 
The quark energy density functional is then given by 
\begin{equation}
    \epsilon_\textrm{q} = 2\int_0^{k_\textrm{F}^\textrm{q}}\frac{\textrm{d}^3k}{(2\pi)^3}\sqrt{M_\textrm{q}^2+k^2}.
\end{equation}

Using this formulation of the quark part to the EOS, it was previously found that the quark onset in this model happens too abruptly, leading to a violation of causality and negative speeds of sound in some cases~\cite{Jeong:2019lhv}. 
Ref.~\cite{Jeong:2019lhv} proposed to cure this issue by scaling the quark density of states by a factor 
\begin{equation}
    g(k) = \frac{\sqrt{\Lambda^2+k^2}}{k}\,,
\end{equation}
where $\Lambda$ is a free parameter. 
This modification is motivated by having the correct high energy behavior, $g(k)\to 1$ as $k\to\infty$, which is consistent with a free Fermi gas. 
The regulator $\Lambda$ can be viewed as the energy scale at which non-perturbative effects become unimportant.
We should, therefore, expect $\Lambda$ to be of the same order of magnitude as the QCD confinement scale $\Lambda_\textrm{QCD}$.

However, Ref.~\cite{Sen:2020qcd} pointed out that this modification to the quark density of states, while curing the unphysical behavior of the speed of sound, introduces a non-negligible density of quarks at all baryon densities, which itself is unphysical. 
Ref.~\cite{Sen:2020qcd} proposed a remedy to this effect by giving a density dependence to $\Lambda$, $\Lambda = \Lambda(n_\textrm{B})$, with
\begin{widetext}
\begin{align}
\Lambda(n_\textrm{B}) = 
\begin{cases}
\Lambda_0\big[-20\big(\frac{n_\textrm{B}}{\nh}\big)^7+70\big(\frac{n_\textrm{B}}{\nh}\big)^6-84\big(\frac{n_\textrm{B}}{\nh}\big)^5+35\big(\frac{n_\textrm{B}}{\nh}\big)^4\big] \quad&\textrm{for}\quad n_\textrm{B}<\nh\,,\\
\Lambda_0 &\textrm{for}\quad n_\textrm{B}\geq \nh\,.\label{lambda(n)}
\end{cases}
\end{align}
\end{widetext}
This choice is motivated as follows: We know that the excluded volume quarkyonic matter model does not produce quarks at low baryon density when the regulator $\Lambda$ is set to zero. 
The need for a nonzero $\Lambda$ arises only to remedy the unphysical behavior of the speed of sound at high densities.
Thus, if we introduce a density dependence in $\Lambda$ such that it is zero at low baryon density and close to the QCD scale near the quark onset density, we will eliminate quarks at low densities while also eliminating the unphysical behavior of the speed of sound at high densities.
If we then want all thermodynamic properties to stay unaffected by $\Lambda$ at $n_\textrm{B}=0$, we need $\Lambda(0) = \textrm{d}^n\Lambda(0)/\textrm{d}(n_\textrm{B})^n = 0$ for $n\leq3$. 
Furthermore, we want $\Lambda(n_\textrm{B})$ to be in full effect once a significant amount of quarks is present. 
It is then natural to impose that $\Lambda(n_\textrm{B}) = \Lambda_0$ for $n_\textrm{B}\geq \nh$, where $\Lambda_0$ is a parameter of order $\Lambda_\textrm{QCD}$. 
We further demand that $\textrm{d}^n\Lambda(n_\textrm{B})/(\textrm{d}n_\textrm{B})^n$ is continuous for $n\leq 3$ to ensure that chemical potential, pressure, and speed of sound are all smooth functions of $n_\textrm{B}$. 
The polynomial \eqref{lambda(n)} was chosen as it is the lowest order polynomial that satisfies all these criteria.  
However, as long as $\Lambda(n_\textrm{B})$ follows the properties discussed above, its exact functional form is not found to be of significant importance~\cite{Sen:2020qcd}.
The quark energy density functional then becomes 
\begin{equation}
    \epsilon_\textrm{q} = 2N_\textrm{c}\int_0^{k_\text{F}^\textrm{q}}\frac{\textrm{d}^3k}{(2\pi)^3}\sqrt{M_\textrm{q}^2+k^2}\frac{\sqrt{\Lambda^2(n_\textrm{B})+k^2}}{k}\,,
\end{equation}
where 
\begin{equation}
    k_\textrm{F}^\textrm{q} = \sqrt{(3\pi^2 n_\textrm{B}^\textrm{q}+\Lambda^3(n_\textrm{B}))^{2/3}-\Lambda^2(n_\textrm{B})}\,.
\end{equation}
We note that the introduction of a density dependent $\Lambda(n_{\text{B}})$ smears out the onset density for the quarks. 
Hence, they appear gradually as a function of baryon density \cite{Sen:2020qcd}. 
However, the quark density is essentially zero compared to the neutron density up to baryon densities close to $\nh$, and so the picture in Fig. \ref{fermi_sphere} still holds.

To allow for two flavors of quarks 
(up and down), we describe the total quark baryon density as 
\begin{equation}
    n_\textrm{B}^\textrm{Q} = n^u_\textrm{B}+n^d_\textrm{B}\,,
\end{equation}
and relate the up ($n^u_\textrm{B}$) and down ($n^d_\textrm{B}$) quark baryon densities by enforcing charge neutrality,
\begin{equation}
    n^d_\textrm{B} = 2n^u_\textrm{B}\,.
\end{equation}
Then, the total quark energy density functional is given by
\begin{equation}
    \epsilon_\textrm{Q} = 2N_\textrm{c}\sum_{q = u,d}\int_0^{k_\text{F}^\textrm{q}}\frac{\textrm{d}^3k}{(2\pi)^3}\sqrt{M_\textrm{q}^2+k^2}\frac{\sqrt{\Lambda^2(n_\textrm{B})+k^2}}{k}\,.\label{e2}
\end{equation}

\subsubsection{Full quarkyonic EOS}

In the quarkyonic model, the free quarks and the neutrons have a single Fermi distribution, where the bottom of the neutron Fermi sea is set by the Fermi momentum of the d-quarks, $k_\textrm{F}^\textrm{d}$. 
That is, $k_\textrm{F} \equiv N_\textrm{c}k_\textrm{F}^\textrm{d}$ as illustrated in Fig.~\ref{fermi_sphere}. 
Combining Eqs.~\eqref{e1} and \eqref{e2}, the total energy density functional is given by
\begin{align}
\epsilon =& 2\Big[1-\Big(\frac{n_\textrm{B}^\textrm{N}}{\nh}\Big)\Big]\int_0^\Delta\frac{\textrm{d}^3k}{(2\pi)^3}\sqrt{M_\textrm{N}^2+k^2}+\tilde{V}(n_\textrm{B}^\textrm{N})
\\
&+2N_\textrm{c}\sum_{q = u,d}\int_0^{k_\text{F}^\textrm{q}}\frac{\textrm{d}^3k}{(2\pi)^3}\sqrt{M_\textrm{q}^2+k^2}\frac{\sqrt{\Lambda^2(n_\textrm{B})+k^2}}{k}\,,\nonumber
\end{align}
where 
\begin{equation}
    \Delta = (3\pi^2 n_\textrm{ex}^\textrm{N}+k_\textrm{F}^3)^{1/3}-k_\textrm{F}\,.
\end{equation}
The equilibrium configuration at a given baryon density is then found by minimizing $\epsilon$ with respect to either $n_\textrm{B}^\textrm{Q}$ or $k_\textrm{F}$. 
Once the equilibrium configuration is found, it is straightforward to obtain the energy density of this equilibrium configuration from the energy density functional. 
Other interesting thermodynamic properties can then be obtained using the thermodynamic relations
\begin{align}
    \mu(n_\textrm{B}) &= \frac{\textrm{d}\epsilon(n_\textrm{B})}{\textrm{d}n_\textrm{B}}\,,\quad\quad \textrm{(chemical potential)}
    \\
    P(n_\textrm{B}) &= n_\textrm{B}\mu(n_\textrm{B})-\epsilon(n_\textrm{B})\,, \quad\textrm{(pressure)}
    \\
    c_S^2(n_\textrm{B}) &= \frac{\textrm{d}P(n_\textrm{B})}{\textrm{d}\epsilon(n_\textrm{B})}\,,\quad\quad\quad \textrm{(speed of sound)}
\end{align}
for the equilibrium configuration.

We would like to point out some of the limitations of the quarkyonic matter model employed here that have not yet been discussed. 
First, we have assumed the baryonic part to be purely made up of neutrons, and have ignored beta equilibrium because it is not trivial to treat the Fermi sphere for quarkyonic matter in a dynamic model when it involves more than one species of nucleons. 
We note, whoever, that there are non-dynamic quarkyonic matter models accounting for beta-equilibrium~\cite{PhysRevC.104.055803,PhysRevD.102.023021}. 
It can nevertheless be argued that the fraction of protons to neutrons is less than $\lesssim 10\%$~\cite{McLerran:2018hbz} and should, therefore, not have a significant effect on neutron star masses and radii. 
Second, we have only considered two flavors of quarks. 
In principle, strange quarks could be included as well and there are three-flavor excluded volume models of quarkyonic matter in the literature~\cite{Duarte:2020xsp,Duarte_2020}. 
However, the inclusion of strange quarks introduces hyperons, for which the nuclear interactions are not well understood. 
This leads to the introduction of additional free parameters, comlicating the analysis in this paper.
Recently, there have been developments in mean-field approaches to quarkyonic matter~\cite{kumar2023quarkyonic,xia2023quarkyonic} as well. 
However, we note that there are no such models that dynamically set the neutron shell width like the excluded volume model we present here. 
We, therefore, leave the analysis of such models for future work. 
The quarkyonic model described in this section was shown to be in agreement with neutron star mass-radius data~\cite{Sen:2020qcd} for $\Lambda_0$ close to the QCD scale and $n_0$ of the order of the nuclear saturation density. 
In the next section, we are going to be agnostic about these and the other model parameters and extensively investigate what astrophysical data tells us about their values, as well as the macroscopic significance of the quarkyonic matter model, using a Bayesian approach.

\subsection{Bayesian Inference}\label{bayesian inference}

By using Bayes' theorem, the posterior $p(\vec{\theta} | d,\mathcal{H})$ on a set of parameters $\vec{\theta}$ under the hypothesis $\mathcal{H}$ and with data $d$ is given by
\begin{equation}
\begin{aligned}
        p(\vec{\theta} | d,\mathcal{H}) =  \frac{p(d|\vec{\theta},\mathcal{H})p(\vec{\theta}|\mathcal{H})}{p(d|\mathcal{H})} \to 
        \mathcal{P}(\vec{\theta}) =  \frac{\mathcal{L}(\vec{\theta})\pi(\vec{\theta})}{\mathcal{Z}}\,,
\end{aligned}
\end{equation}
where $\mathcal{P}(\vec{\theta})$, $\mathcal{L}(\vec{\theta})$, $\pi(\vec{\theta})$, and $\mathcal{Z}$ are the posterior, likelihood, prior, and evidence, respectively. 
The prior describes our knowledge of the parameters before any observations. The likelihood quantifies how likely the hypothesis can describe the data at a given point $\vec{\theta}$ in the parameter space.
Finally, the evidence, also known as the marginalized likelihood, marginalizes the likelihood over the whole parameter space with respect to the prior, i.e,
\begin{equation}
    \mathcal{Z} = \int_{\mathcal{V}} d\vec{\theta} \mathcal{L}(\vec{\theta})\pi(\vec{\theta})\,.
\end{equation}

Two hypotheses, $\mathcal{H}_1$ and $\mathcal{H}_2$, can be compared by calculating the odds ratio between them, $\mathcal{O}^1_2$, given by
\begin{equation}
    \mathcal{O}^1_2 \equiv \frac{p(d|\mathcal{H}_1)}{p(d|\mathcal{H}_2)}\frac{p(\mathcal{H}_1)}{p(\mathcal{H}_2)} \equiv \mathcal{B}^1_2\Pi^1_2\,,
\end{equation}
where $\mathcal{B}^1_2$ and $\Pi^1_2$ are the Bayes factor and prior odds, respectively. 
If $\mathcal{O}^1_2 > 1$, $\mathcal{H}_1$ is more plausible than $\mathcal{H}_2$, and vice versa. Throughout this work, the prior odds is set to $1$, in which case the Bayes factor is the same as the odd ratio.

To combine multiple independent observations, we express the likelihood as
\begin{equation}
    \mathcal{L}(\vec{\theta}) = \prod_i \mathcal{L}_i(\vec{\theta})\,,
\end{equation}
where $\mathcal{L}_i(\vec{\theta})$ is the likelihood corresponding to the $i$-th observation. Therefore, the combined posterior $\mathcal{P}(\vec{\theta})$ is given by
\begin{equation}
\begin{aligned}
    \mathcal{P}(\vec{\theta}) &= \frac{\pi(\vec{\theta})}{\mathcal{Z}}\mathcal{L}(\vec{\theta}) = \frac{\pi(\vec{\theta})}{\mathcal{Z}}\prod_i \mathcal{L}_i(\vec{\theta})\,.
\end{aligned}
\end{equation}

\subsection{Implementation}\label{implementation}

The Nuclear-Physics Multi-Messenger Astrophysics framework NMMA~\cite{Dietrich:2020lps,Pang:2022rzc} has been known for its capability and flexibility of including various multi-messenger astrophysical observations, nuclear-theory calculations, and heavy-ion collision experiments~\cite{Huth:2021bsp}. 
Here, however, we only consider the astrophysical observations to constrain our theoretical model so that we obtain an astrophysics-only motivated result.

\begin{figure}
    \centering
    \includegraphics[width=\columnwidth]{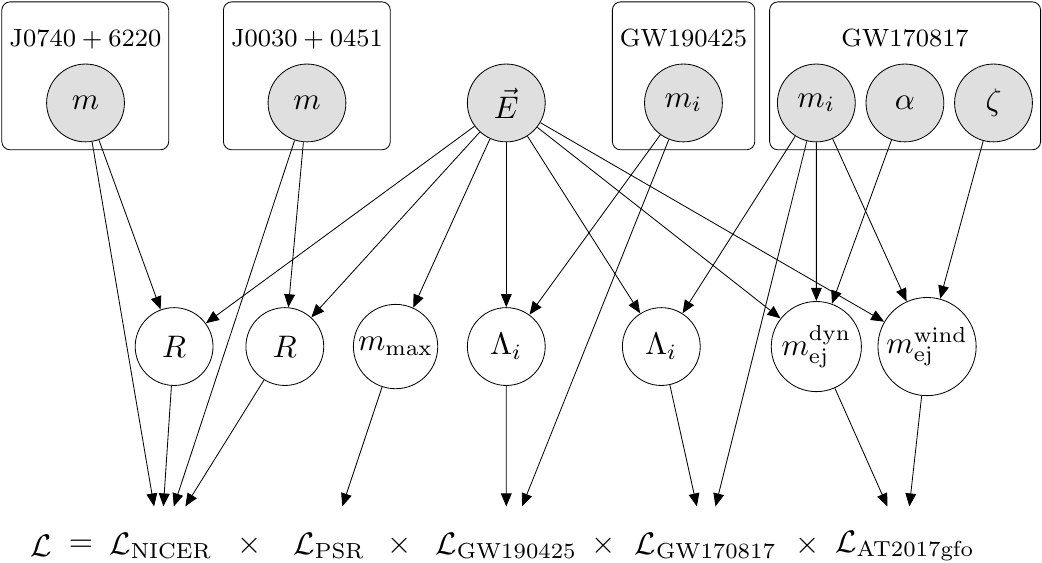}
    \caption{The inference logic of our multi-messenger analysis. Each shaded node corresponds to parameter(s) to be sampled. The unshaded nodes refer to the latent parameters calculated based on the sample parameters and required for the likelihood evaluation. An arrow pointing from $A$ to $B$ refers to $B$ depending on $A$. The NICER likelihood $\mathcal{L}_{\rm NICER} \equiv \mathcal{L}_{\rm J0740+6620} \times \mathcal{L}_{\rm J0030+0451}$, and the PSR likelihood $\mathcal{L}_{\rm PSR} \equiv \mathcal{L}_{\rm J0348+4032} \times \mathcal{L}_{\rm J1614-2230}$. In total, eight neutron stars are taken into account for the analysis.}
    \label{fig:inference_relation}
\end{figure}

To calculate the likelihood for different observational data given the quarkyonic EOS model, a similar approach as in previous works is used~\cite{Lackey:2014fwa,Essick:2020flb,Wysocki:2020myz,Pang:2020ilf}. 
In particular, the likelihood of the EOS parameters $\vec{E}\equiv\{\tilde{a}, \tilde{b}, \nh, \Lambda_0\}$ given the $i$-th observation is given by
\begin{equation}
\begin{aligned}
        \mathcal{L}(\vec{E}) = \int d\vec{\theta}_{\small\textrm{macro}} \frac{\pi(\vec{\theta}_{\small\textrm{macro}}|\vec{E})}{\pi(\vec{\theta}_{\small\textrm{macro}})}\mathcal{P}(\vec{\theta}_{\small\textrm{macro}})\,,
\end{aligned}
\end{equation}
where $\vec{\theta}_{\small\textrm{macro}}$, $\pi(\vec{\theta}_{\small\textrm{macro}}|\vec{E})$, and $\pi(\vec{\theta}_{\small\textrm{macro}})$ are the macroscopic parameters of interest, and the priors on $\vec{\theta}_{\small\textrm{macro}}$ with and without an EOS imposed, respectively. 
The inference logic is summarized in Fig.~\ref{fig:inference_relation}. The joint posterior is explored using the Nested Sampling algorithm Multinest~\cite{Feroz_2009} implemented in \texttt{PyMultinest}~\cite{pymultinest1,pymultinest2}. 
The details of the likelihood evaluation for each observation are described in the following. 

\subsubsection{PSR J0348+4032 and PSR J1614-2230}

The radio observations on PSR~J0348+4042~\cite{Antoniadis:2013pzd}, and PSR~J1614-2230~\cite{Arzoumanian:2017puf} have provided a lower bound on the maximum mass of a neutron star. 
For these two observations, the mass of the pulsar is the macroscopic parameter of interest.
The likelihood is given by
\begin{equation}
    \mathcal{L}_{{\rm PSR-}j}(\vec{E}) = \int^{m_{\rm max}}_{0}dm \mathcal{P}(m | {\rm PSR-}j)\,,
\end{equation}
where $\mathcal{P}(m | {\rm PSR-}j)$ is the posterior distribution of the pulsar $j$'s mass, and $m_{\rm max}$ is the maximum mass supported by the EOS with parameter $\vec{E}$.
We approximate the posterior distribution of the two pulsars to be Gaussian with the reported values and $1$-$\sigma$ uncertainty, similar to Ref.~\cite{Dietrich:2020lps}. 

\subsubsection{PSR J0030+0451 and PSR J0740+6620}

The Neutron Star Interior Composition Explorer (NICER) mission aims to measure both the masses and the radii of pulsars.
It has provided estimates for the mass and radius for the pulsars J0030+0451~\cite{Miller:2019cac,Riley:2019yda} and J0740+6620~\cite{Miller:2021qha,Riley:2021pdl}\footnote{An update of the NICER analysis on J0740+6620 is presented in Ref.~\cite{Salmi:2022cgy}. Due to the agreement between the previous results and the updated one, here we are using the previously original results.}. 
For J0740+6620, data from the XMM-Newton telescope~\cite{Struder:2001bh, Turner:2000jy} has been additionally used for improving the total flux measurement.

The corresponding likelihood for the NICER measurement of pulsar $j$ is given by
\begin{equation}
\begin{aligned}
    &\mathcal{L}_{{\rm NICER-}j}(\vec{E}, m)\\
    &= \int d\!R\ \mathcal{P}_{{\rm NICER}-j}(m, R)\frac{\pi(m, R |\vec{E})}{\pi(m, R | I)}\\
	&\propto \int d\!R\ \mathcal{P}_{{\rm NICER}-j}(m, R)\delta(R-R(m,\vec{E}))\\
	&\propto \mathcal{P}_{{\rm NICER}-j}(m, R=R(m, \vec{E}))\,,
\end{aligned}
\end{equation}
where $\mathcal{P}_{{\rm NICER}-j}(m, R)$ is the joint-posterior distribution of mass and radius as measured by NICER and we use the fact that i) the radius is a function of mass for a given EOS, and ii) the prior for the mass and radius is taken to be uniform in Refs.~\cite{Miller:2019cac,Riley:2019yda,Miller:2021qha,Riley:2021pdl}.\footnote{The mass measurement for J0740+6620 using radio observations~\cite{Fonseca2021} is taken as the prior on the mass for the NICER analysis on J0740+6620~\cite{Miller:2021qha,Riley:2021pdl}. 
Hence, we consider it to be part of the NICER likelihood for that star. 
Therefore, the prior on the mass of J0740+6620 can be considered uniform.}
Similar to Ref.~\cite{Huth:2021bsp}, we use the results of Ref.~\cite{Miller:2019cac} for PSR~J0030+0451,  while for PSR J0740+6620 we average over the results presented in Refs.~\cite{Miller:2021qha} and~\cite{Riley:2021pdl}.

\subsubsection{GW170817 and GW190425}

By analyzing the gravitational-wave signals GW170817~\cite{TheLIGOScientific:2017qsa} and GW190425~\cite{Abbott:2020uma}, the masses $m_i$ and the tidal deformability $\Lambda_i$ of the two neutron stars in the binary can be estimated\footnote{Note that it was also proposed that GW190425 originated from a neutron-star-black-hole merger~\cite{Foley:2020kus, Han:2020qmn, Kyutoku:2020xka}.}. 
The corresponding likelihood is given by
\begin{equation}
\begin{aligned}
        \mathcal{L}_{{\rm GW}-j}(\vec{E}, m_i) &= \int d\Lambda_i \frac{\pi(m_i,\Lambda_i|\vec{E})}{\pi(m_i,\Lambda_i|I)} \mathcal{P}_{{\rm GW}-j}(m_i,\Lambda_i)\\
        &= \int d\Lambda_i \frac{\prod_i\delta(\Lambda_i-\Lambda(m_i; \vec{E}))}{\pi(\Lambda_i| m_i, I)}\\
        &\times\frac{\pi(m_i|\vec{E})}{\pi(m_i|I)} \mathcal{P}_{{\rm GW}-j}(m_i,\Lambda_i)\\
        &\propto \left.\frac{\mathcal{P}_{{\rm GW}-j}(m_i,\Lambda_i)}{\pi(\Lambda_i|, m_i, I)}\right\vert_{\Lambda_i=\Lambda(m_i;\vec{E})}\,,
\end{aligned}
\end{equation}
where $\mathcal{P}_{{\rm GW}-j}(m_i,\Lambda_i)$ is the joint-posterior distribution on mass and tidal deformability of the binary neutron star $j$ measured by its gravitational-wave signal.
We use that the tidal deformability is a function of mass for a given EOS.
For both events, we use the publicly available posterior samples~\cite{GW170817_PE_samples,GW190425_PE_samples}.

\subsubsection{AT2017gfo}
\label{sec:at2017gfo}
For analyzing the observed kilonova AT2017gfo, we have done the Bayesian inference with a Gaussian likelihood function given by
\begin{equation}
    \mathcal{L}_{\rm EM}(\vec{\theta}) \propto \exp \left(-\frac{1}{2}\sum_{ij}\left(\frac{m^{j}_{i} - m^{j, \rm{est}}_{i}(\vec{\theta})}{\sigma^{j}_{i}}\right)^2\right)\,,
\end{equation}
where $m^{j}_{i}$ and $\sigma^{j}_{i}$ are the observed apparent magnitude and its corresponding statistical uncertainties, at observation time $t_i$, respectively. 
Moreover, $m^{j, \rm{est}}_{i}$ are the theoretically predicted apparent magnitudes for a given filter $j$ (a passband for a particular wavelength interval).

For this analysis, the model presented in Ref.~\cite{Bulla:2019muo} is used. For this model, the macroscopic parameters of interest are the dynamic ejecta mass $m^{\rm dyn}_{\rm ej}$ and the wind ejecta mass $m^{\rm wind}_{\rm ej}$. 
The dynamic ejecta refer to the material ejected during the merger via torque and shocks while the wind ejecta refer to material released from the disk formed during the merger.

In order to connect the ejecta masses with the EOS and the masses of the two neutron stars in the binary, fits to numerical-relativity simulations are used. 
For the dynamical ejecta mass $m^{\rm dyn}_{\rm ej}$, the fit formula is given by~\cite{Kruger:2020gig}
\begin{equation}
    \frac{m^{\rm dyn}_{\rm ej, fit}}{10^{-3}M_{\odot}} = \left(\frac{a}{C_1} + b\left(\frac{m_2}{m_1}\right)^n + c\ C_1\right) + (1 \leftrightarrow 2)\,,
\end{equation}
where $m_i$ and $C_i$ are the masses and the compactness of the two components of the binary and the best-fit coefficients are given by $a=-9.3335$, $b=114.17$, $c=-337.56$, and $n=1.5465$. 
As presented in Ref.~\cite{Kruger:2020gig}, this relation provides an accurate estimation of the ejecta mass with an error that is well-approximated by a zero-mean Gaussian with a standard deviation of $0.004M_{\odot}$.
Therefore, the dynamic ejecta mass is described by
\begin{equation}
    m^{\rm dyn}_{\rm ej} = m^{\rm dyn}_{\rm ej, fit} + \alpha\,,
    \label{dyn_ejecta_mass}
\end{equation}
where $\alpha\sim\mathcal{N}(\mu=0, \sigma=0.004M_{\odot})$.

For the wind ejecta mass, we assume it to be proportional to the disk mass,
\begin{equation}
    m^{\rm wind}_{\rm ej} = \zeta \times m_{\rm disk, fit}\,,
\end{equation}
where $\zeta$ is an independent parameter in $[0,1)$. 
To estimate the disk mass $m_{\rm disk}$, we follow the relation presented in Ref.~\cite{Dietrich:2020lps},
\begin{align}
    &\log_{10}\left(\frac{m_{\rm disk, fit}}{M_{\odot}}\right) =\label{eq:Mdisk_fit} \\
    &\textrm{max}\left(-3, a\left(1+b\tanh\left(\frac{c - (m_1+m_2) M_{\rm thr}^{-1}}{d}\right)\right)\right)\,, \nonumber
\end{align}
with $a$ and $b$ given by
\begin{equation}
\begin{aligned}
    a &= a_o + \delta a \cdot \xi\,,\\
    b &= b_o + \delta b \cdot \xi\,.
\end{aligned}
\end{equation}
The parameter $\xi$ is given by
\begin{equation}
    \xi = \frac{1}{2}\tanh\left(\beta \left(q-q_{\rm trans}\right)\right)\,,
\end{equation}
where $q \equiv m_2/m_1 \leq 1$ is the mass ratio and $\beta$ and $q_{\rm trans}$ are free parameters. 
The best-fitting parameters are $a_o=-1.725$, $\delta a=-2.337$, $b_o=-0.564$, $\delta b=-0.437$, $c =0.958$, $d=0.057$, $\beta=5.879$, $q_{\rm trans}=0.886$. 
The threshold mass $M_{\rm thr}$ is estimated to follow the prediction presented in Ref.~\cite{Bauswein:2020aag}. 
Because a fitting error in the disk mass is degenerate with the proportionality parameters $\zeta$, no additional error parameters, similar to $\alpha$ introduced in Eq.~\eqref{dyn_ejecta_mass}, are included. 
We point out that the relations Eqs.~\eqref{dyn_ejecta_mass} and \eqref{eq:Mdisk_fit} are generally calibrated to a large set of numerical-relativity simulations covering a large range of the parameter space, and while these relations are found to be `quasi-universal', i.e., applicable independent of the exact EOS, they have not been tested against quarkyonic binary neutron stars mergers, which could result in a hidden systematic uncertainty in our analysis. 

\begin{table}[]
    \centering
    \begin{tabular}{|c | l|}
         \hline
         Parameter & Prior\\
         \hline
         $\tilde{a}$ & Uniform$(-50{\rm MeV}, 50{\rm MeV})$\\
         $\tilde{b}$ & Uniform$(-50{\rm MeV}, 50{\rm MeV})$\\
         $\nh$ & LogUniform$(10^{-2}n_{\rm sat}, 20n_{\rm sat})$\\
         $\Lambda_0$ & LogUniform$(10^{-5}{\rm GeV}, 10{\rm GeV})$\\
         \hline
    \end{tabular}
    \caption{Priors imposed for the EOS parameters in the analysis. 
    ``Uniform$(a,b)$'' is a uniform distribution in the range of $[a, b)$ and ``LogUniform$(a,b)$'' is a log-uniform distribution in the range of $[a, b)$. 
    Therefore, if $x \sim {\rm LogUniform}(a, b)$, $\log x \sim {\rm Uniform}(\log a, \log b)$.}
    \label{tab:prior}
\end{table}

The likelihood for the EOS parameters $\vec{E}$ and the component masses of the two neutron stars, $m_i$ is given by
\begin{align}
    &\mathcal{L}_{\rm AT2017gfo}(\vec{E}, m_i, \alpha, \zeta)\\
    &=\int dm^{\rm dyn}_{\rm ej}dm^{\rm wind}_{\rm ej} \frac{\pi(m^{\rm dyn}_{\rm ej}, m^{\rm wind}_{\rm ej} | \vec{E}, m_i, \alpha, \zeta)}{\pi(m^{\rm dyn}_{\rm ej}, m^{\rm wind}_{\rm ej} | I)}\nonumber\\
    &\times \mathcal{P}_{\rm AT2017gfo}(m^{\rm dyn}_{\rm ej}, m^{\rm wind}_{\rm ej})\nonumber\\
    &\propto \mathcal{P}_{\rm AT2017gfo}(m^{\rm dyn}_{\rm ej}, m^{\rm wind}_{\rm ej})\vert_{m^{\rm dyn}_{\rm ej} = m^{\rm dyn}_{\rm ej}(\vec{E}, m_i, \alpha), m^{\rm wind}_{\rm ej}(\vec{E}, m_i, \zeta)}\,. \nonumber
\end{align}

\section{Results}\label{sec:results}

\begin{figure*}[t]
    \centering
    \includegraphics[width=0.7\textwidth]{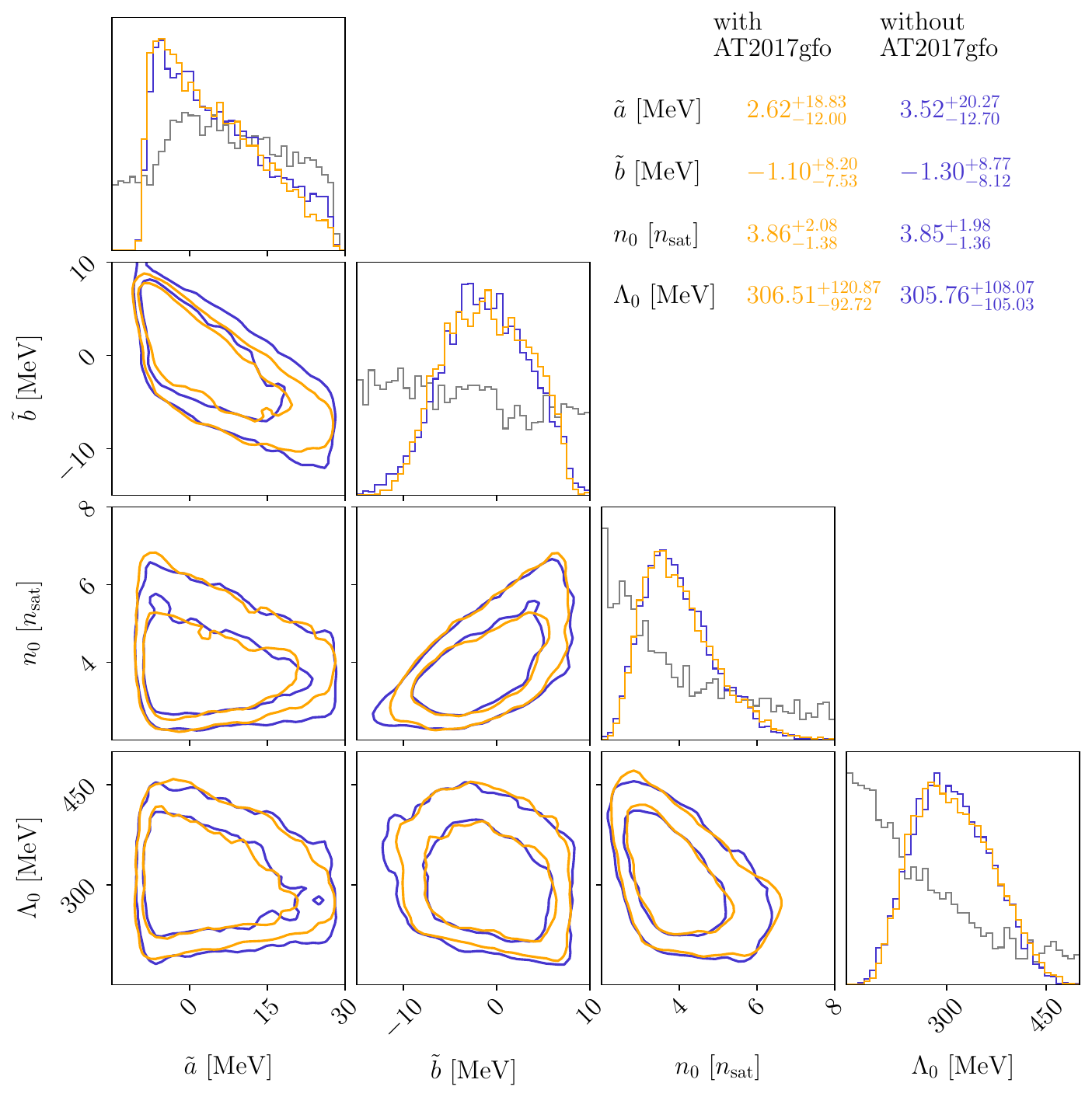}
    \caption{Posterior on the microscopic parameters of the quarkyonic EOS model. The 2D contours are shown at $68\%$ and $95\%$ levels. The priors on each of the parameters are also shown (grey). In addition to the distributions shown in Tab.~\ref{tab:prior}, the priors shown additionally impose that the resulting EOS can support a neutron star.}
    \label{fig:microparameters}
\end{figure*}

In the following section, we discuss the results of our analysis of astrophysical multi-messenger observations when applying the quarkyonic EOS model.
As mentioned in Sec.~\ref{sec:at2017gfo}, the `quasi-universal' relations needed for analyzing AT2017gfo have not been tested against the quarkyonic model. Therefore, we report results without the inclusion of AT2017gfo as our main, conservative results and results with the inclusion of AT2017gfo in parentheses.
All quoted values are medians with their $95\%$ credible intervals unless mentioned otherwise.

\subsection{Estimation on EOS parameters}

The priors for the EOS parameters $\tilde{a}$, $\tilde{b}$, $\nh$, and $\Lambda_0$ are defined Tab.~\ref{tab:prior}.
Additionally, we impose that
\begin{itemize}
    \item each parameter sample results in a valid EOS, i.e.,
    \begin{itemize}
        \item the crust and core EOSs intercept, and
        \item the EOS explores densities beyond the neutron-star crust,
    \end{itemize}
    \item and the EOS produces stable neutron stars, indicated by a stable $M$-$R$-$\Lambda$ curve.
\end{itemize}
In Fig.~\ref{fig:microparameters}, we show the priors and the posteriors with their median values and with the uncertainty quoted at $95\%$ credibility. 

\begin{figure}[t]
    \centering
    \includegraphics[width=\linewidth]{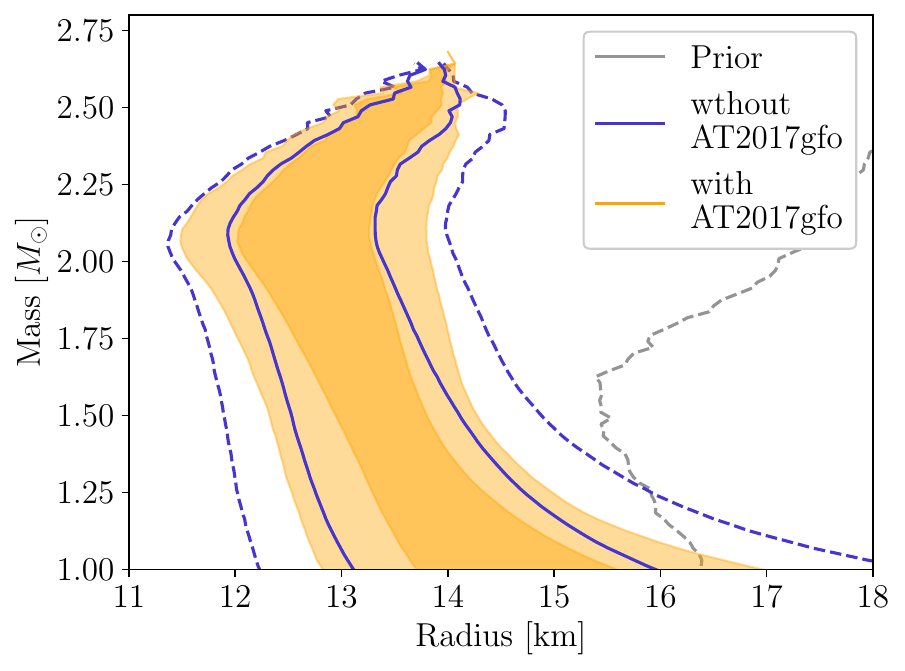}
    \caption{Posterior on the neutron star radius as a function of mass. The darker/solid (lighter/dashed) band shows the 68\% (95\%) credible interval of the radius at a given mass. The prior band is shown in the 95\% band.}
    \label{fig:mass_radius}
\end{figure}

The posterior on $\tilde{a}$ skews towards positive values, while the posterior on $\tilde{b}$ is centered around $0$, with both of them having sizeable uncertainty. This shows that interactions with higher-order density dependences between neutrons are not pronounced within this model. 
This finding can be attributed to the relatively low $\nh$, which indicates the characteristic density at which quark matter starts to appear.
This low $\nh$ causes the quark part of the EOS to dominate over the neutrons' interactions.
However, we note that there are large uncertainties in the neutron matter parameters $\tilde{a}$ and $\tilde{b}$, mainly due to the fact that we have only used astrophysical observations to constrain them.

More interestingly, we find that the parameter $\nh$ is favored to be on the order of a few times the nuclear saturation density, which is comparable to the densities reached inside the heaviest neutron stars. 
This implies that, assuming our quarkyonic matter model describes supranuclear-dense matter, there will be a significant amount of quarks in the core of neutron stars with masses in the two-solar-mass range. 
We will investigate this in more detail in the next section.

Finally, we comment on the parameter $\Lambda_0$. 
As mentioned in section \ref{free_quarks}, the inclusion of $\Lambda_0$ was motivated by regulating the infrared divergence that occurs at low densities due to the non-perturbative effects of QCD. 
We, therefore, expect that $\Lambda_0\approx\Lambda_{\text{QCD}}$. 
If we take $\Lambda_{\text{QCD}}$ to be defined by the Landau pole of the QCD strong coupling, one finds $\Lambda_{\text{QCD}}\approx 200-400$MeV, depending on the renormalization scheme used~\cite{PhysRevD.92.054008}. 
This is indeed consistent with what we obtain for $\Lambda_0$, as seen in Fig.~\ref{fig:microparameters}. 
We emphasize that although this is an expected result, the value of $\Lambda_0$ is extracted completely by comparison with astrophysical data. 
It is interesting and reassuring that observations of neutron stars can constrain this microscopic parameter of our model in a way that is consistent with expectations from QCD. 
This is because the confinement scale is important for the high-density behavior of the EOS, which, in turn, has a high impact on the mass of a neutron star. 
Conversely, this explains, in part, why we are not able to constrain $\tilde{a}$ and $\tilde{b}$ very well, as they mostly describe the low-density behavior of the EOS. 
These findings are consistent with similar results showing that the correlation between the low- and high-density EOS can be broken in neutron stars~\cite{Essick:2021kjb,Essick:2021ezp}.

The posterior of the radius of neutron stars as a function of mass is shown in Fig.~\ref{fig:mass_radius}. 
Within this quarkyonic-matter model, the radius of a $1.4M_\odot$ neutron star, $R_{1.4}$, is estimated to be $13.44^{+1.69}_{-1.54} (13.54^{+1.02}_{-1.04})$ km, at 95\% credibility, without (with) the inclusion of AT2017gfo; the kilonova observation significantly tightens the radius constraint. 
The extracted value for $R_{1.4}$ is higher than in previous estimations (e.g. Tab. 1 in Ref.~\cite{Pang:2022rzc})\footnote{Please note that the NICER collaboration recently published a re-analysis of their observation of J0030+0451\cite{Vinciguerra:2023qxq}, which could influence the inferred properties of a canonical neutron star.}, reflecting the difference of model-independent and model-dependent data analyses of neutron-star observations, as well as the absence of the constraint from chiral effective field theory at low densities~\cite{Tews:2018kmu}.
The latter also contributes to the wide posteriors on $\tilde{a}$ and $\tilde{b}$. 

\subsection{Presence of quark matter}

\begin{figure}
    \centering
    \includegraphics[width=\linewidth]{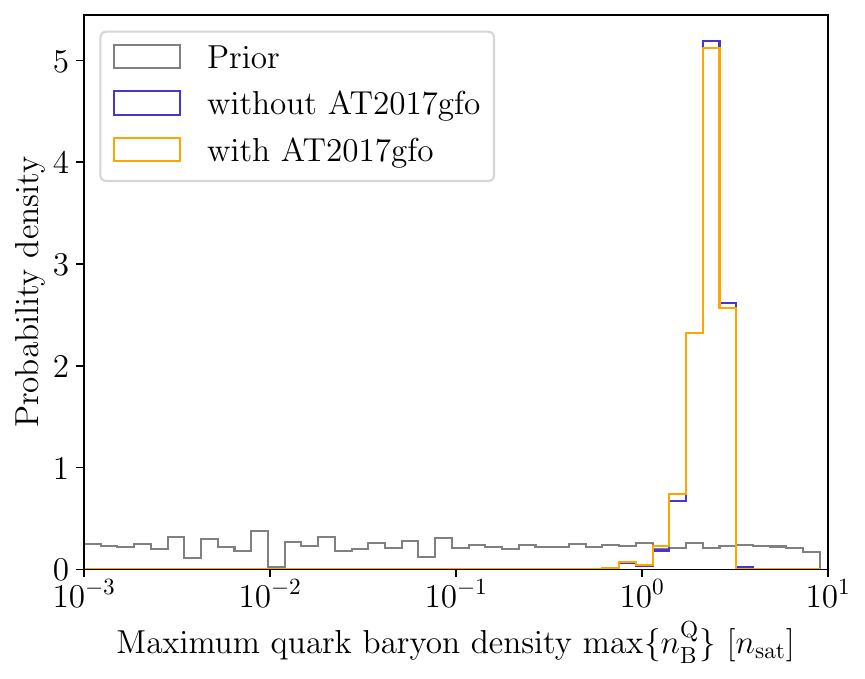}
    \caption{Posterior on the maximum quark density observed within a neutron star. The posterior and prior are re-weighted such that the prior on the maximum quark density is log-uniform, i.e., uniform in $\log n^{\rm Q}_{\rm B}$.}
    \label{fig:max_quark_density}
\end{figure}

\begin{figure}
    \centering
    \includegraphics[width=\linewidth]{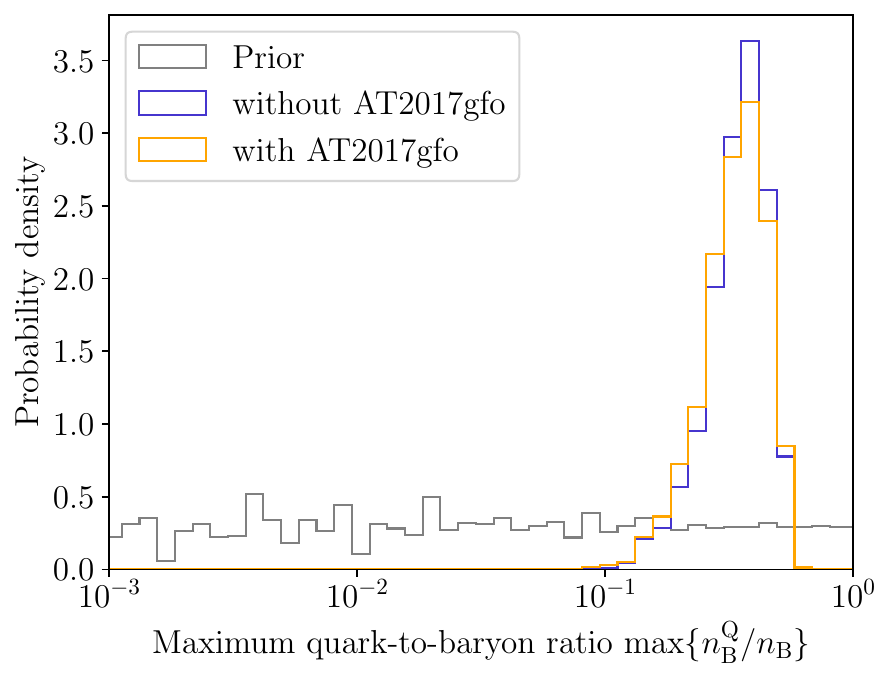}
    \caption{Posterior on the maximum quark-to-baryon ratio observed within a neutron star. The posterior and prior are re-weighted such that the prior on the ratio is log-uniform, i.e., uniform in $\log n^{\rm Q}_{\rm B}/n_{\rm B}$.}
    \label{fig:max_quark_to_baryon_density}
\end{figure}

\begin{figure*}
    \centering
    \includegraphics[width=0.7\linewidth]{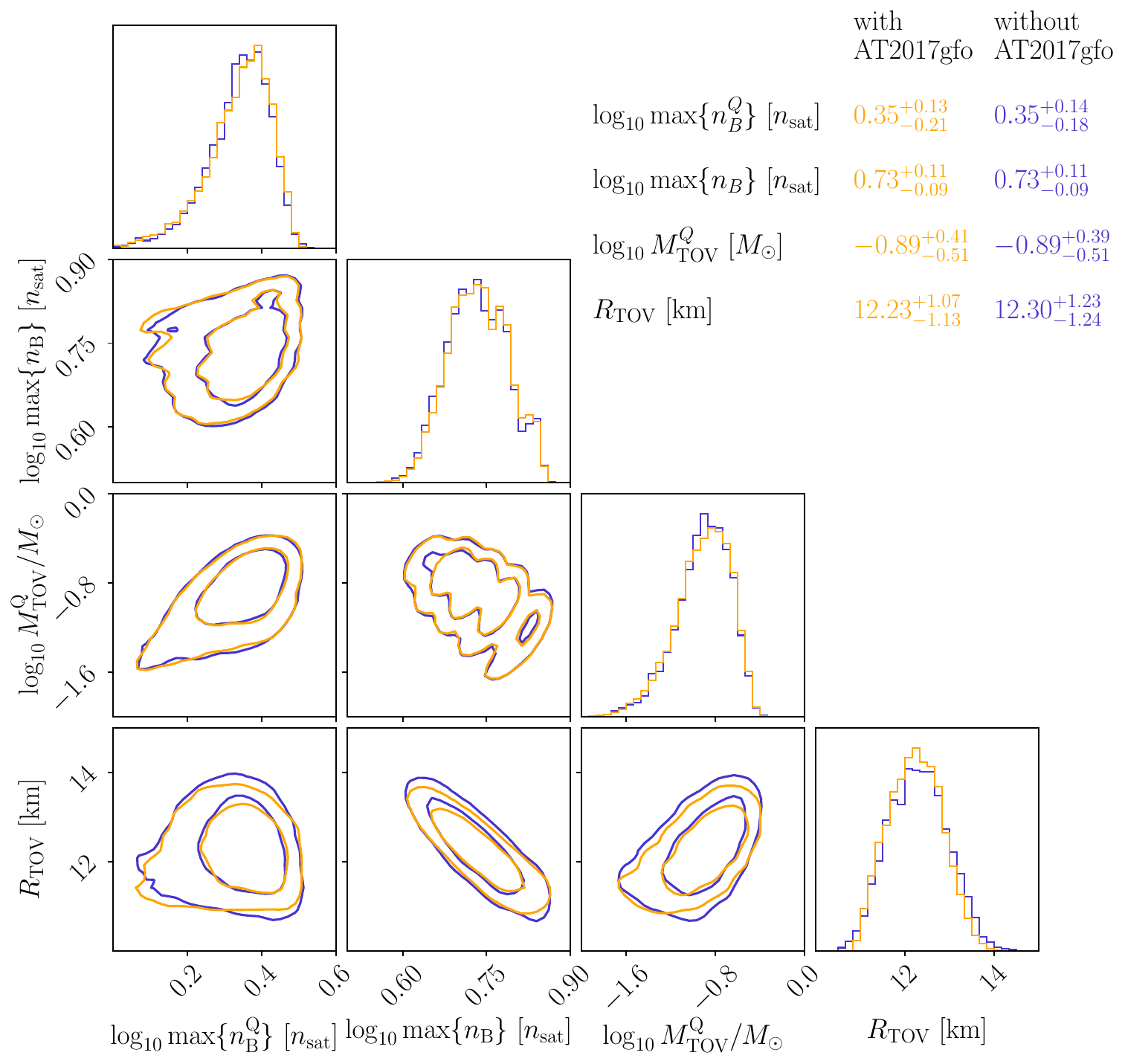}
    \caption{Posterior on the ($\log_{10}$) maximum quark baryon density $\max\{n^{\rm Q}_{\rm B}\}$, the ($\log_{10}$) maximum baryon density $\max\{n_{\rm B}\}$, the ($\log_{10}$) quark core mass for maximal mass neutron star $M^{\rm Q}_{\rm TOV}$ and the associated radius $R_{\rm TOV}$. A strong positive correlation is observed between the quark core mass and the radius, and between maximum quark baryon density and quark core mass.}
    \label{fig:rTOV_corner}
\end{figure*}

To study the presence of quarks in the core of neutron stars, we estimate the maximum quark baryon density $\max\{n^{\rm Q}_{\rm B}\}$ and the maximum quark-to-baryon ratio $\max\{n^{\rm Q}_{\rm B} / n_{\rm B}\}$. 
That is, we consider the density of quarks and quark-to-baryon ratio inside the center of the neutron star with largest mass. 
The estimated distributions of $\max\{n^{\rm Q}_{\rm B}\}$ and $\max\{n^{\rm Q}_{\rm B} / n_{\rm B}\}$ are shown in Fig.~\ref{fig:max_quark_density} and Fig.~\ref{fig:max_quark_to_baryon_density}, respectively. 
In both cases, we re-weighted the distributions to respect a log-uniform prior to enhance the impact of astrophysical observations. 
Based on the estimated distributions, the quarkyonic model is predicting $\max\{n^{\rm Q}_{\rm B}\} \gtrsim 1 n_{\rm sat}$ and the ratio $\max\{n^{\rm Q}_{\rm B} / n_{\rm B}\} \gtrsim 10\%$, showing a significant presence of quarks within the most massive neutron stars.

\begin{table}[]
    \centering
    \renewcommand{\arraystretch}{1.2}
    \begin{tabular}{|l|c|c|c|c|}
    \hline
     & \multicolumn{2}{c|}{Without AT2017gfo} & \multicolumn{2}{c|}{With AT2017gfo}\\
     \hline
     & $m_q [M_\odot]$ & $m_q / m [\%]$  & $m_q[M_\odot]$ & $m_q / m [\%]$\\
    \hline
    GW170817-$m_1$ & $0.01^{+0.02}_{-0.01}$ & $0.41^{+1.45}_{-0.37}$ & $0.00^{+0.02}_{-0.00}$ & $0.34^{+1.36}_{-0.31}$ \\
    GW170817-$m_2$ & $0.00^{+0.01}_{-0.00}$ & $0.23^{+0.91}_{-0.20}$ & $0.00^{+0.01}_{-0.00}$ & $0.25^{+1.07}_{-0.22}$\\
    GW190425-$m_1$ & $0.02^{+0.05}_{-0.02}$ & $0.98^{+2.88}_{-0.94}$ & $0.02^{+0.06}_{-0.02}$ & $0.98^{+3.16}_{-0.94}$\\
    GW190425-$m_2$ & $0.01^{+0.03}_{-0.01}$ & $0.52^{+1.79}_{-0.48}$ & $0.01^{+0.03}_{-0.01}$ & $0.51^{+1.96}_{-0.47}$\\
    PSR J0030+0451 & $0.01^{+0.03}_{-0.01}$ & $0.40^{+1.65}_{-0.36}$ & $0.01^{+0.03}_{-0.01}$ & $0.40^{+1.74}_{-0.37}$\\
    PSR J0740+6620 & $0.05^{+0.11}_{-0.05}$ & $2.57^{+5.41}_{-2.48}$ & $0.05^{+0.12}_{-0.05}$ & $2.57^{+5.82}_{-2.47}$\\
    Maximal mass & $0.13^{+0.14}_{-0.10}$ & $5.87^{+6.72}_{-4.72}$& $0.13^{+0.14}_{-0.11}$ & $5.86^{+7.08}_{-4.87}$\\
    \hline
    \end{tabular}
    \caption{Estimated quark core mass $m_q$ within neutron stars and the associated mass fraction $m_q / m$ for the stars.}
    \label{tab:quark_core_mass}
\end{table}

To give a concrete picture of the abundance of quarks in neutron stars, we further estimate the total mass that the free quarks contribute to neutron stars, see Tab.~\ref{tab:quark_core_mass}. 
As expected, the quark core mass is low for most of the neutron stars. 
However, PSR J0740+6620, which is one of the most massive neutron stars observed, would carry about $\approx 0.05M_\odot$ of free quarks, or  $\approx2.6\%$ of its mass. 
For a neutron star at maximum mass, the quark core can grow to $\approx 0.13M_\odot$, or $\approx 5.9\%$ of the stellar mass.

Moreover, a strong presence of free quarks shows a correlation with macroscopic parameters. In Fig.~\ref{fig:rTOV_corner}, we show a corner plot connecting the maximum quark baryon density $\max\{n^{\rm Q}_{\rm B}\}$, the maximum baryon density $\max\{n_{\rm B}\}$, the quark mass in the maximal-mass neutron star $M^{\rm Q}_{\rm TOV}$ and the associated radius $R_{\rm TOV}$. 
We observe a strong positive correlation between the quark core mass and the radius for the maximal mass neutron star within the quarkyonic-matter model. Although it may seem counter-intuitive that stars containing quarks have large radii, we note that a star's radius is mostly attributed to matter properties at intermediate densities. At these densities, the parameter $n_0$ roughly sets the density at which quarks appear, leading to a sudden increase in the pressure gradient around $n_B=n_0$. The earlier this increase happens, the larger the radii of the star will be. A larger radii is then consistent with having more quarks, considering that the amount of quarks inside a star is inversely correlated with the $n_0$ parameter.  
\subsection{Speed of sound and normalized trace anomaly}

\begin{figure}
    \centering
    \includegraphics[width=\linewidth]{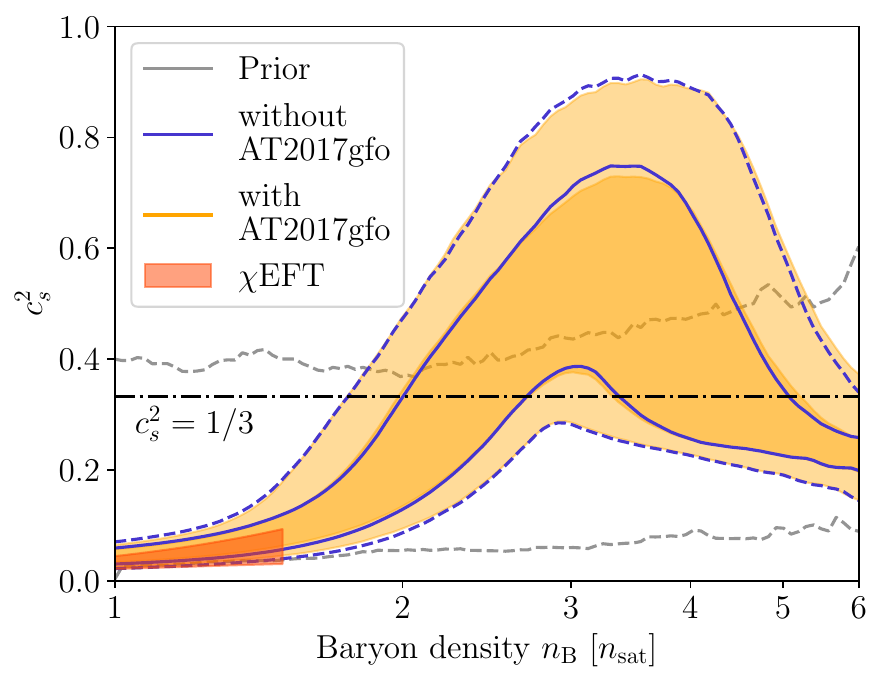}
    \caption{Posterior on the speed of sound $c^2_s$ as a function of density. The darker/solid (lighter/dashed) band show the $68\%$ ($95\%$) credible interval of the speed of sound at a given density. The prior band is shown at  $95\%$ confidence level. 
    Moreover, we show the range predicted from the chiral effective field theory calculation of Ref.~\cite{Tews:2018kmu} up to $1.5n_{\rm sat}$.}
    \label{fig:c2_over_n}
\end{figure}

\begin{figure}[t]
    \centering
    \includegraphics[width=\linewidth]{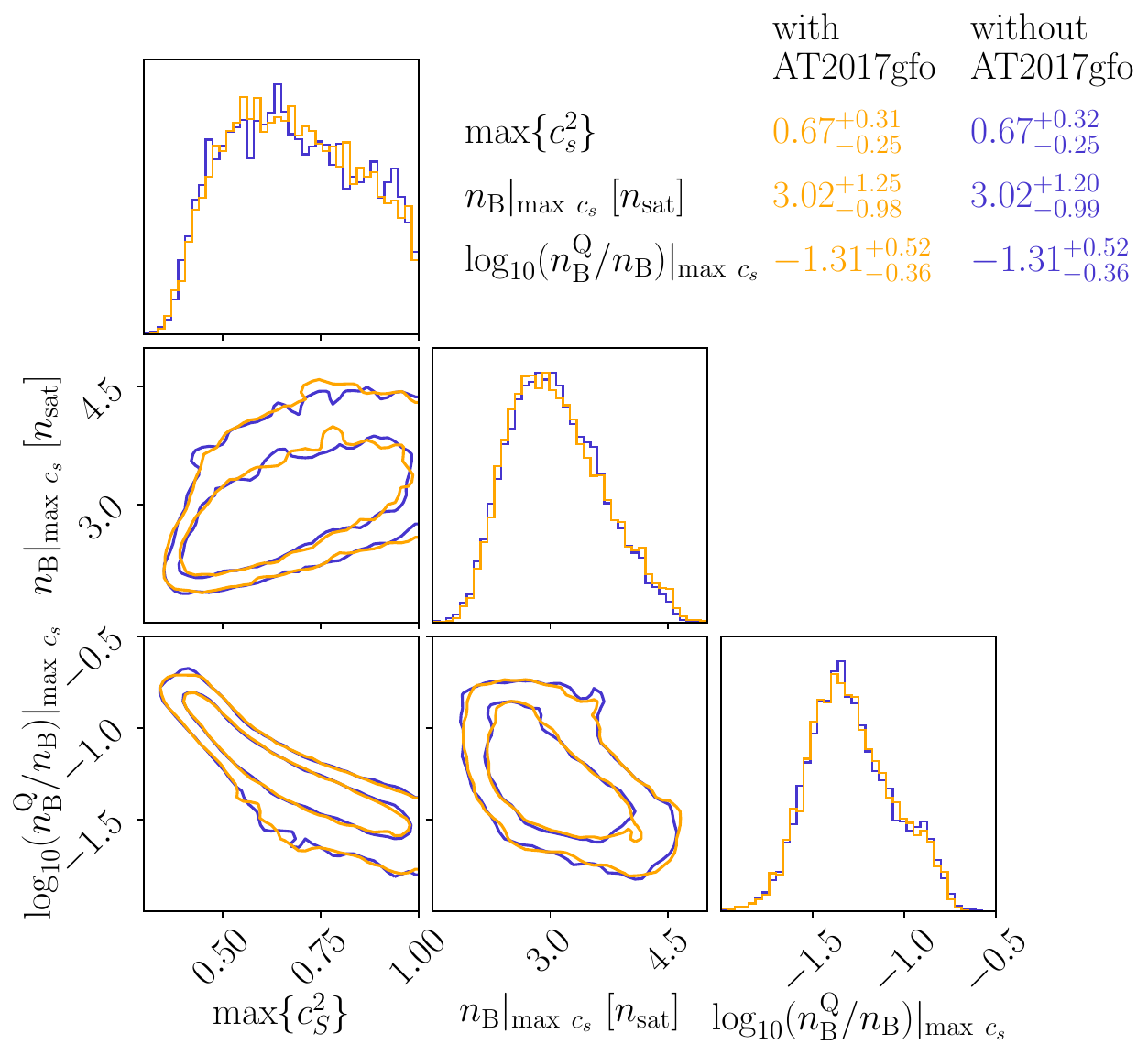}
    \caption{Posterior on the maximum speed of sound $\max\{c^2_s\}$ and the associated number density $n_{\rm B}\vert_{{\rm max} \ c_s}$ and the quark-to-baryon ratio $n^{\rm Q}_{\rm B} / n_{\rm B}\vert_{{\rm max} \ c_s}$. A strong negative correlation is observed between the maximum speed of sound and the ratio $n^{\rm Q}_{\rm B} / n_{\rm B}\vert_{{\rm max} \ c_s}$.}
    \label{fig:max_c2_corner}
\end{figure}

We further examine the speed of sound  within neutron stars and show it as a function of density in Fig.~\ref{fig:c2_over_n}. 
It is obvious that the quarkyonic-matter model violates the conformal limit of $c^2_s = 1/3$ at about 3 times the nuclear saturation density, but the speed of sound drops below that limit at higher densities.
For comparison, the range predicted by chiral effective field theory is also shown in Fig.~\ref{fig:c2_over_n}, and is on the lower end of the quarkyonic-matter prediction. 
The maximum speed-of-sound, the densities at which the maximum occurs, and the quark-to-baryon ratio at this density are shown in the corner plot in Fig.~\ref{fig:max_c2_corner}. 
The estimate for the maximum speed of sound and its density agree with the result reported in Ref.~\cite{Legred:2021hdx}, in which a nonparametric EOS model is used. 
The maximum speed of sound and the quark-to-baryon ratio show a strong negative correlation, reflecting the intuition that quark matter is generally softer than nucleonic matter.

\begin{figure}[t]
    \centering
    \includegraphics[width=\linewidth]{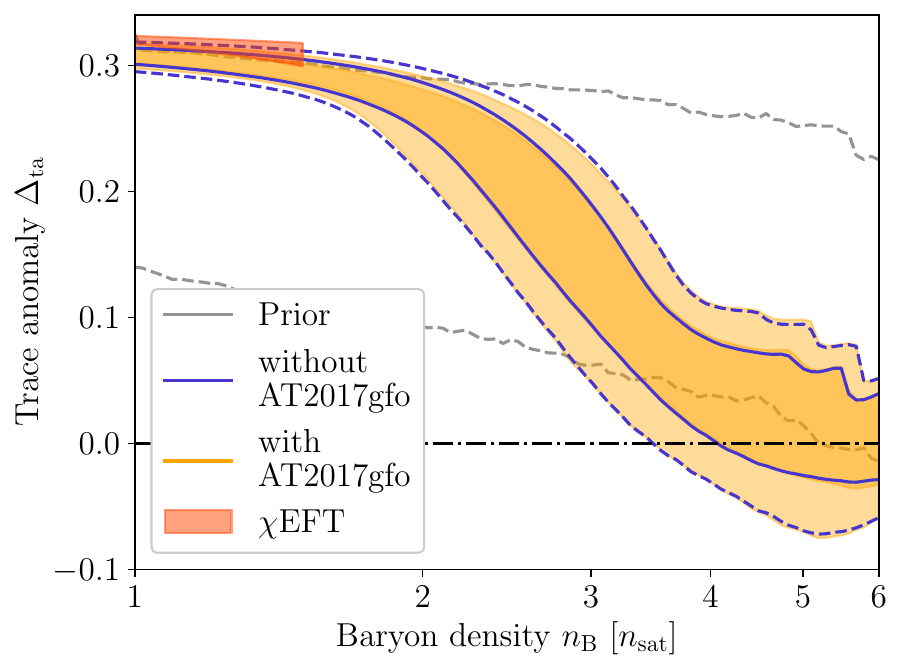}
    \caption{Posterior on the normalized trace anomaly $\Delta_{\rm ta}$ as a function of density. The darker/solid (lighter/dashed) band show the $68\%$ ($95\%$) credible interval of the speed of sound at a given density. The prior band is shown in the $95\%$ band. Moreover, the predicted range based on the chiral effective field theory calculation of Ref.~\cite{Tews:2018kmu} is shown up to $1.5n_{\rm sat}$.}
    \label{fig:Delta_over_n}
\end{figure}

Recently, the authors of Ref.~\cite{Fujimoto:2022ohj} suggested that a peak in the speed of sound, as in Fig.~\ref{fig:c2_over_n}, is a sign of strongly-coupled conformal matter.
This was demonstrated by examining the behaviour of the normalized trace anomaly, which is defined as
\begin{equation}
    \Delta_{\rm ta} \equiv \frac{1}{3} - \frac{P}{\epsilon}\,.
\end{equation}
From the definition it is evident that, as matter approaches conformality, $\Delta_{\rm ta} \to 0$. 
Even though the speed of sound and the trace anomaly are closely related, the behavior at intermediate densities is different and provides complementary information. 
By expressing the sound speed in terms of the trace anomaly and its derivative, Ref.~\cite{Fujimoto:2022ohj} showed that the peak in the sound speed is, in fact, not a violation of the conformal bound, but it is a steep approach to the conformal limit. 
In this work, we check this interpretation by computing the trace anomaly as a function of the density, see Fig.~\ref{fig:Delta_over_n}, and comparing it with the behavior of the sound speed. 
In Fig.~\ref{fig:Delta_over_n}, we show that at $n \approx 6n_{\rm sat}$, $\Delta_{\rm ta} \to 0$, as can be inferred from the behavior of the speed of sound in Fig.~\ref{fig:c2_over_n}. 
However, at intermediate densities, $\Delta_{\rm ta}$ decreases towards its conformal value, exactly at those densities where we find a peak in the sound speed. 
These results are similar to what was observed in Ref.~\cite{Fujimoto:2022ohj}. 
Note, however, that the quarkyonic-matter model employed in this work does not explicitly include any interactions between quarks and therefore it cannot be interpreted as strongly coupled matter. 
This suggests that inferring the presence of strongly coupled conformal matter from neutron-star observations is non-trivial since the corresponding characteristic behaviour of the trace anomaly can also be produced by weakly interacting matter.

Given the presence of quarks within the neutron stars within the quarkyonic-matter model, we further investigate the consistency between the model and the perturbative QCD (pQCD) calculations of the EOS at $\approx 40 n_{\rm sat}$~\cite{Gorda:2021znl}. 
Following the approach of Ref.~\cite{Somasundaram:2022ztm}, we connect to the integral pQCD constraints of Ref.~\cite{Komoltsev:2021jzg} at the maximum-mass configuration because we are interested only in the EOS of stable neutron stars.\footnote{See Ref.~\cite{Gorda:2022jvk} for a similar analysis but with a matching density of $10 n_{\rm sat}$.}
We use the parameters associated with the maximally massive neutron star, i.e., the associated central pressure $p_{\rm TOV}$, the baryon number density $n_{\rm TOV}$ and the chemical potential $\mu_{\rm TOV}$ to check if a given EOS is consistent with the high-density pQCD constraint by verifying
\begin{enumerate}
    \item $\Delta p_{\rm min} < p_{\rm pQCD} - p_{\rm TOV} < \Delta p_{\rm max}$, and,
    \item $n_{\rm TOV} < n_{\rm pQCD}$,
\end{enumerate}
where $p_{\rm pQCD}$ and $n_{\rm pQCD}$ are functions of $\mu_{\rm pQCD}$, which is set to $2.6{\rm GeV}$. 
Furthermore, $\Delta p_{\rm min}$ and $\Delta p_{\rm max}$ are calculated using Eqs.~(12) and~(13) of Ref.~\cite{Gorda:2022jvk} respectively. 
If the proposed EOS satisfies both of the above criteria, it has a consistency of $100\%$ and $0\%$ otherwise.
Since the pQCD calculation~\cite{Gorda:2021znl} depends on the renormalization scale parameter $X$, we calculate the consistency of a given EOS with pQCD by marginalizing over $X \sim {\rm LogUniform}(1, 4)$. 
Given the posterior samples, without (with) the inclusion of AT2017gfo, the average consistency, i.e., the average of the aforementioned calculation over all EOSs, is $99.88\%$ ($99.89\%$) with a minimum consistency of $93.9\%$ ($94.7\%$). 
Therefore, we conclude that the quarkyonic-matter model is consistent with pQCD calculations.

\section{Conclusions}\label{sec:conclusion}

In this work, we have studied a dynamic quarkyonic-matter model for supranuclear matter inside neutron stars. 
This model is employed within the nuclear multi-messenger astrophysics framework, which includes radio and X-ray (NICER) observations of pulsars, gravitational-wave observations from binary neutron star mergers, in particular GW170817, and electromagnetic observations of the kilonova associated with GW170817. 
Assuming that the quarkyonic-matter model describes neutron stars, we constrain the properties of quarkyonic matter using astrophysical observations. 
The main results can be summarized as follows:
\begin{itemize}
    \item We constrained the four  model parameters without (with) AT2017gfo at 95\% credibility to be 
    \begin{align}
        \tilde{a} &= 2.62^{+18.83}_{-12.00}(3.52^{+20.27}_{-12.70})\text{MeV},
        \nonumber\\ \tilde{b}&=-1.10^{+8.20}_{-7.53}(-1.30^{+8.77}_{-8.12})\text{MeV},
        \nonumber
        \\
        \nh &= 3.86^{+2.08}_{-1.38}(3.85^{1.98}_{-1.36})n_{\text{sat}},
        \nonumber\\ \Lambda_0&=306.51^{+120.87}_{-92.72}(305.76^{+108.07}_{-105.03})\text{MeV}.
    \end{align}
    \item The model is predicting a substantial presence of quarks within neutron stars, in particular:
    \begin{itemize}
        \item The maximum quark baryon density exceeds $1n_{\rm sat}$,
        \item The maximum quark-to-baryon ratio exceeds $10\%$\, and 
        \item The mass contribution due to quarks can reach $\approx0.13M_\odot$, attributing to $\approx 6\%$ of the neutron star's mass.
    \end{itemize}
    \item The quarkyonic-matter model predicts a peak in the speed-of-sound in neutron stars, and the conformal limit is subsequently restored at $\approx 6n_{\rm sat}$.
    \item The quarkyonic-matter model predicts $R_{1.4}=13.44^{+1.69}_{-1.54} (13.54^{+1.02}_{-1.04})$~km at 95\% credibility level, without (with) the inclusion of AT2017gfo.
\end{itemize}

Our work demonstrates that the quarkyonic-matter model can explain the current observations on neutron stars and allows for a significant presence of quarks within them. It is, therefore, a viable alternative to other models, including free quarks.

Moreover, we have shown that the observations of neutron stars constrain the nuclear equation-of-state at higher densities, while the constraints at lower densities are weaker.
At these densities, nuclear experiments such as FOPI~\cite{LeFevre:2015paj}, ASY-EOS~\cite{Russotto:2016ucm}, or PREX~\cite{PREXII}, and nuclear theory calculation, e.g., in the framework of  $\chi$EFT, have a larger impact on determinations of the equation-of-state. Therefore, for a complete picture of the nuclear equation of state such constraints should be included. 
In addition, recent observation of a low-mass pulsar, HESS J1731-347~\cite{Doroshenko2022}, can also shed light on the constraint on nuclear matter at a lower density.

Although it is impossible to establish the existence of quarkyonic matter with current neutron-star observations, future observations using next-generation gravitational-wave observatories, such as the Einstein telescope~\cite{Punturo:2010zza, Hild:2010id} or Cosmic Explorer~\cite{Reitze:2019iox}, will likely provide crucial new data. 
Together with advancements in theoretical nuclear physics, this will allow us to better understand the nature of dense matter in the future.

\acknowledgements

P.T.H.P. and C.V.D.B. are supported by the research programme of the Netherlands Organisation for Scientific Research (NWO). L.S. and S.S. are supported by the U.S. Department of Energy, Nuclear Physics Quantum Horizons program through the Early Career Award No. DE-SC0021892. R.S. acknowledges support from the Nuclear Physics from Multi-Messenger Mergers (NP3M) Focused Research Hub which is funded by the National Science Foundation under Grant No. 21-16686, and by the Laboratory Directed Research and Development program of Los Alamos National Laboratory under Project No. 20220541ECR. T.D. acknowledges funding from the EU Horizon under ERC Starting Grant, No. SMArt-101076369, support from the Deutsche Forschungsgemeinschaft, DFG, Project No. DI 2553/7, and from the Daimler and Benz Foundation for the project “NUMANJI”. Views and opinions expressed are those of the authors only and do not necessarily reflect those of the European Union or the Eu ropean Research Council. Neither the European Union nor the granting authority can be held responsible for them. The work of I.T. was supported by the U.S. Department of Energy, Office of Science, Office of Nuclear Physics, under Contract No. DE-AC52-06NA25396, by the Laboratory Directed Research and Development program of Los Alamos National Laboratory under Projects No. 20220541ECR and No. 20230315ER, and by the U.S. Department of Energy, Office of Science, Office of Advanced Scientific Computing Research, Scientific Discovery through Advanced Computing (SciDAC) NUCLEI program. This material is based upon work supported by NSF's LIGO Laboratory which is a major facility fully funded by the National Science Foundation. M.W.C acknowledges support from the National Science Foundation with Grants No. PHY-2308862 and No. PHY-2117997.

\bibliography{refs.bib}

\begin{thebibliography}{77}%
\makeatletter
\providecommand \@ifxundefined [1]{%
 \@ifx{#1\undefined}
}%
\providecommand \@ifnum [1]{%
 \ifnum #1\expandafter \@firstoftwo
 \else \expandafter \@secondoftwo
 \fi
}%
\providecommand \@ifx [1]{%
 \ifx #1\expandafter \@firstoftwo
 \else \expandafter \@secondoftwo
 \fi
}%
\providecommand \natexlab [1]{#1}%
\providecommand \enquote  [1]{``#1''}%
\providecommand \bibnamefont  [1]{#1}%
\providecommand \bibfnamefont [1]{#1}%
\providecommand \citenamefont [1]{#1}%
\providecommand \href@noop [0]{\@secondoftwo}%
\providecommand \href [0]{\begingroup \@sanitize@url \@href}%
\providecommand \@href[1]{\@@startlink{#1}\@@href}%
\providecommand \@@href[1]{\endgroup#1\@@endlink}%
\providecommand \@sanitize@url [0]{\catcode `\\12\catcode `\$12\catcode
  `\&12\catcode `\#12\catcode `\^12\catcode `\_12\catcode `\%12\relax}%
\providecommand \@@startlink[1]{}%
\providecommand \@@endlink[0]{}%
\providecommand \url  [0]{\begingroup\@sanitize@url \@url }%
\providecommand \@url [1]{\endgroup\@href {#1}{\urlprefix }}%
\providecommand \urlprefix  [0]{URL }%
\providecommand \Eprint [0]{\href }%
\providecommand \doibase [0]{http://dx.doi.org/}%
\providecommand \selectlanguage [0]{\@gobble}%
\providecommand \bibinfo  [0]{\@secondoftwo}%
\providecommand \bibfield  [0]{\@secondoftwo}%
\providecommand \translation [1]{[#1]}%
\providecommand \BibitemOpen [0]{}%
\providecommand \bibitemStop [0]{}%
\providecommand \bibitemNoStop [0]{.\EOS\space}%
\providecommand \EOS [0]{\spacefactor3000\relax}%
\providecommand \BibitemShut  [1]{\csname bibitem#1\endcsname}%
\let\auto@bib@innerbib\@empty
\bibitem [{\citenamefont {Ozel}\ and\ \citenamefont
  {Freire}(2016)}]{ozel:2016oaf}%
  \BibitemOpen
  \bibfield  {author} {\bibinfo {author} {\bibfnamefont {F.}~\bibnamefont
  {Ozel}}\ and\ \bibinfo {author} {\bibfnamefont {P.}~\bibnamefont {Freire}},\
  }\href {\doibase 10.1146/annurev-astro-081915-023322} {\bibfield  {journal}
  {\bibinfo  {journal} {Ann. Rev. Astron. Astrophys.}\ }\textbf {\bibinfo
  {volume} {54}},\ \bibinfo {pages} {401} (\bibinfo {year} {2016})},\ \Eprint
  {http://arxiv.org/abs/1603.02698} {arXiv:1603.02698 [astro-ph.HE]}
  \BibitemShut {NoStop}%
\bibitem [{\citenamefont {Flanagan}\ and\ \citenamefont
  {Hinderer}(2008)}]{Flanagan:2007ix}%
  \BibitemOpen
  \bibfield  {author} {\bibinfo {author} {\bibfnamefont {E.~E.}\ \bibnamefont
  {Flanagan}}\ and\ \bibinfo {author} {\bibfnamefont {T.}~\bibnamefont
  {Hinderer}},\ }\href {\doibase 10.1103/PhysRevD.77.021502} {\bibfield
  {journal} {\bibinfo  {journal} {Phys.Rev.}\ }\textbf {\bibinfo {volume}
  {D77}},\ \bibinfo {pages} {021502} (\bibinfo {year} {2008})},\ \Eprint
  {http://arxiv.org/abs/0709.1915} {arXiv:0709.1915 [astro-ph]} \BibitemShut
  {NoStop}%
\bibitem [{\citenamefont {Del~Pozzo}\ \emph {et~al.}(2013)\citenamefont
  {Del~Pozzo}, \citenamefont {Li}, \citenamefont {Agathos}, \citenamefont {Van
  Den~Broeck},\ and\ \citenamefont {Vitale}}]{DelPozzo:2013ala}%
  \BibitemOpen
  \bibfield  {author} {\bibinfo {author} {\bibfnamefont {W.}~\bibnamefont
  {Del~Pozzo}}, \bibinfo {author} {\bibfnamefont {T.~G.~F.}\ \bibnamefont
  {Li}}, \bibinfo {author} {\bibfnamefont {M.}~\bibnamefont {Agathos}},
  \bibinfo {author} {\bibfnamefont {C.}~\bibnamefont {Van Den~Broeck}}, \ and\
  \bibinfo {author} {\bibfnamefont {S.}~\bibnamefont {Vitale}},\ }\href
  {\doibase 10.1103/PhysRevLett.111.071101} {\bibfield  {journal} {\bibinfo
  {journal} {Phys. Rev. Lett.}\ }\textbf {\bibinfo {volume} {111}},\ \bibinfo
  {pages} {071101} (\bibinfo {year} {2013})},\ \Eprint
  {http://arxiv.org/abs/1307.8338} {arXiv:1307.8338 [gr-qc]} \BibitemShut
  {NoStop}%
\bibitem [{\citenamefont {Annala}\ \emph {et~al.}(2018)\citenamefont {Annala},
  \citenamefont {Gorda}, \citenamefont {Kurkela},\ and\ \citenamefont
  {Vuorinen}}]{Annala:2017llu}%
  \BibitemOpen
  \bibfield  {author} {\bibinfo {author} {\bibfnamefont {E.}~\bibnamefont
  {Annala}}, \bibinfo {author} {\bibfnamefont {T.}~\bibnamefont {Gorda}},
  \bibinfo {author} {\bibfnamefont {A.}~\bibnamefont {Kurkela}}, \ and\
  \bibinfo {author} {\bibfnamefont {A.}~\bibnamefont {Vuorinen}},\ }\href
  {\doibase 10.1103/PhysRevLett.120.172703} {\bibfield  {journal} {\bibinfo
  {journal} {Phys. Rev. Lett.}\ }\textbf {\bibinfo {volume} {120}},\ \bibinfo
  {pages} {172703} (\bibinfo {year} {2018})},\ \Eprint
  {http://arxiv.org/abs/1711.02644} {arXiv:1711.02644 [astro-ph.HE]}
  \BibitemShut {NoStop}%
\bibitem [{\citenamefont {Abbott}\ \emph {et~al.}(2018)\citenamefont {Abbott}
  \emph {et~al.}}]{Abbott:2018exr}%
  \BibitemOpen
  \bibfield  {author} {\bibinfo {author} {\bibfnamefont {B.~P.}\ \bibnamefont
  {Abbott}} \emph {et~al.} (\bibinfo {collaboration} {Virgo, LIGO
  Scientific}),\ }\href {\doibase 10.1103/PhysRevLett.121.161101} {\bibfield
  {journal} {\bibinfo  {journal} {Phys. Rev. Lett.}\ }\textbf {\bibinfo
  {volume} {121}},\ \bibinfo {pages} {161101} (\bibinfo {year} {2018})},\
  \Eprint {http://arxiv.org/abs/1805.11581} {arXiv:1805.11581 [gr-qc]}
  \BibitemShut {NoStop}%
\bibitem [{\citenamefont {De}\ \emph {et~al.}(2018)\citenamefont {De},
  \citenamefont {Finstad}, \citenamefont {Lattimer}, \citenamefont {Brown},
  \citenamefont {Berger},\ and\ \citenamefont {Biwer}}]{De:2018uhw}%
  \BibitemOpen
  \bibfield  {author} {\bibinfo {author} {\bibfnamefont {S.}~\bibnamefont
  {De}}, \bibinfo {author} {\bibfnamefont {D.}~\bibnamefont {Finstad}},
  \bibinfo {author} {\bibfnamefont {J.~M.}\ \bibnamefont {Lattimer}}, \bibinfo
  {author} {\bibfnamefont {D.~A.}\ \bibnamefont {Brown}}, \bibinfo {author}
  {\bibfnamefont {E.}~\bibnamefont {Berger}}, \ and\ \bibinfo {author}
  {\bibfnamefont {C.~M.}\ \bibnamefont {Biwer}},\ }\href {\doibase
  10.1103/PhysRevLett.121.091102} {\bibfield  {journal} {\bibinfo  {journal}
  {Phys. Rev. Lett.}\ }\textbf {\bibinfo {volume} {121}},\ \bibinfo {pages}
  {091102} (\bibinfo {year} {2018})},\ \Eprint
  {http://arxiv.org/abs/1804.08583} {arXiv:1804.08583 [astro-ph.HE]}
  \BibitemShut {NoStop}%
\bibitem [{\citenamefont {Tews}\ \emph
  {et~al.}(2018{\natexlab{a}})\citenamefont {Tews}, \citenamefont {Margueron},\
  and\ \citenamefont {Reddy}}]{Tews:2018iwm}%
  \BibitemOpen
  \bibfield  {author} {\bibinfo {author} {\bibfnamefont {I.}~\bibnamefont
  {Tews}}, \bibinfo {author} {\bibfnamefont {J.}~\bibnamefont {Margueron}}, \
  and\ \bibinfo {author} {\bibfnamefont {S.}~\bibnamefont {Reddy}},\ }\href
  {\doibase 10.1103/PhysRevC.98.045804} {\bibfield  {journal} {\bibinfo
  {journal} {Phys. Rev. C}\ }\textbf {\bibinfo {volume} {98}},\ \bibinfo
  {pages} {045804} (\bibinfo {year} {2018}{\natexlab{a}})},\ \Eprint
  {http://arxiv.org/abs/1804.02783} {arXiv:1804.02783 [nucl-th]} \BibitemShut
  {NoStop}%
\bibitem [{\citenamefont {Watts}\ \emph {et~al.}(2016)\citenamefont {Watts}
  \emph {et~al.}}]{Watts:2016uzu}%
  \BibitemOpen
  \bibfield  {author} {\bibinfo {author} {\bibfnamefont {A.~L.}\ \bibnamefont
  {Watts}} \emph {et~al.},\ }\href {\doibase 10.1103/RevModPhys.88.021001}
  {\bibfield  {journal} {\bibinfo  {journal} {Rev. Mod. Phys.}\ }\textbf
  {\bibinfo {volume} {88}},\ \bibinfo {pages} {021001} (\bibinfo {year}
  {2016})},\ \Eprint {http://arxiv.org/abs/1602.01081} {arXiv:1602.01081
  [astro-ph.HE]} \BibitemShut {NoStop}%
\bibitem [{\citenamefont {Demorest}\ \emph {et~al.}(2010)\citenamefont
  {Demorest}, \citenamefont {Pennucci}, \citenamefont {Ransom}, \citenamefont
  {Roberts},\ and\ \citenamefont {Hessels}}]{Demorest:2010bx}%
  \BibitemOpen
  \bibfield  {author} {\bibinfo {author} {\bibfnamefont {P.}~\bibnamefont
  {Demorest}}, \bibinfo {author} {\bibfnamefont {T.}~\bibnamefont {Pennucci}},
  \bibinfo {author} {\bibfnamefont {S.}~\bibnamefont {Ransom}}, \bibinfo
  {author} {\bibfnamefont {M.}~\bibnamefont {Roberts}}, \ and\ \bibinfo
  {author} {\bibfnamefont {J.}~\bibnamefont {Hessels}},\ }\href {\doibase
  10.1038/nature09466} {\bibfield  {journal} {\bibinfo  {journal} {Nature}\
  }\textbf {\bibinfo {volume} {467}},\ \bibinfo {pages} {1081} (\bibinfo {year}
  {2010})},\ \Eprint {http://arxiv.org/abs/1010.5788} {arXiv:1010.5788
  [astro-ph.HE]} \BibitemShut {NoStop}%
\bibitem [{\citenamefont {Antoniadis}\ \emph {et~al.}(2013)\citenamefont
  {Antoniadis}, \citenamefont {Freire}, \citenamefont {Wex}, \citenamefont
  {Tauris}, \citenamefont {Lynch} \emph {et~al.}}]{Antoniadis:2013pzd}%
  \BibitemOpen
  \bibfield  {author} {\bibinfo {author} {\bibfnamefont {J.}~\bibnamefont
  {Antoniadis}}, \bibinfo {author} {\bibfnamefont {P.~C.}\ \bibnamefont
  {Freire}}, \bibinfo {author} {\bibfnamefont {N.}~\bibnamefont {Wex}},
  \bibinfo {author} {\bibfnamefont {T.~M.}\ \bibnamefont {Tauris}}, \bibinfo
  {author} {\bibfnamefont {R.~S.}\ \bibnamefont {Lynch}},  \emph {et~al.},\
  }\href {\doibase 10.1126/science.1233232} {\bibfield  {journal} {\bibinfo
  {journal} {Science}\ }\textbf {\bibinfo {volume} {340}},\ \bibinfo {pages}
  {6131} (\bibinfo {year} {2013})},\ \Eprint {http://arxiv.org/abs/1304.6875}
  {arXiv:1304.6875 [astro-ph.HE]} \BibitemShut {NoStop}%
\bibitem [{\citenamefont {Drischler}\ \emph {et~al.}(2021)\citenamefont
  {Drischler}, \citenamefont {Han}, \citenamefont {Lattimer}, \citenamefont
  {Prakash}, \citenamefont {Reddy},\ and\ \citenamefont
  {Zhao}}]{PhysRevC.103.045808}%
  \BibitemOpen
  \bibfield  {author} {\bibinfo {author} {\bibfnamefont {C.}~\bibnamefont
  {Drischler}}, \bibinfo {author} {\bibfnamefont {S.}~\bibnamefont {Han}},
  \bibinfo {author} {\bibfnamefont {J.~M.}\ \bibnamefont {Lattimer}}, \bibinfo
  {author} {\bibfnamefont {M.}~\bibnamefont {Prakash}}, \bibinfo {author}
  {\bibfnamefont {S.}~\bibnamefont {Reddy}}, \ and\ \bibinfo {author}
  {\bibfnamefont {T.}~\bibnamefont {Zhao}},\ }\href {\doibase
  10.1103/PhysRevC.103.045808} {\bibfield  {journal} {\bibinfo  {journal}
  {Phys. Rev. C}\ }\textbf {\bibinfo {volume} {103}},\ \bibinfo {pages}
  {045808} (\bibinfo {year} {2021})}\BibitemShut {NoStop}%
\bibitem [{\citenamefont {Raaijmakers}\ \emph {et~al.}(2020)\citenamefont
  {Raaijmakers}, \citenamefont {Greif}, \citenamefont {Riley}, \citenamefont
  {Hinderer}, \citenamefont {Hebeler}, \citenamefont {Schwenk}, \citenamefont
  {Watts}, \citenamefont {Nissanke}, \citenamefont {Guillot}, \citenamefont
  {Lattimer},\ and\ \citenamefont {Ludlam}}]{Raaijmakers_2020}%
  \BibitemOpen
  \bibfield  {author} {\bibinfo {author} {\bibfnamefont {G.}~\bibnamefont
  {Raaijmakers}}, \bibinfo {author} {\bibfnamefont {S.~K.}\ \bibnamefont
  {Greif}}, \bibinfo {author} {\bibfnamefont {T.~E.}\ \bibnamefont {Riley}},
  \bibinfo {author} {\bibfnamefont {T.}~\bibnamefont {Hinderer}}, \bibinfo
  {author} {\bibfnamefont {K.}~\bibnamefont {Hebeler}}, \bibinfo {author}
  {\bibfnamefont {A.}~\bibnamefont {Schwenk}}, \bibinfo {author} {\bibfnamefont
  {A.~L.}\ \bibnamefont {Watts}}, \bibinfo {author} {\bibfnamefont
  {S.}~\bibnamefont {Nissanke}}, \bibinfo {author} {\bibfnamefont
  {S.}~\bibnamefont {Guillot}}, \bibinfo {author} {\bibfnamefont {J.~M.}\
  \bibnamefont {Lattimer}}, \ and\ \bibinfo {author} {\bibfnamefont {R.~M.}\
  \bibnamefont {Ludlam}},\ }\href {\doibase 10.3847/2041-8213/ab822f}
  {\bibfield  {journal} {\bibinfo  {journal} {The Astrophysical Journal}\
  }\textbf {\bibinfo {volume} {893}},\ \bibinfo {pages} {L21} (\bibinfo {year}
  {2020})}\BibitemShut {NoStop}%
\bibitem [{\citenamefont {Hebeler}\ \emph {et~al.}(2013)\citenamefont
  {Hebeler}, \citenamefont {Lattimer}, \citenamefont {Pethick},\ and\
  \citenamefont {Schwenk}}]{Hebeler:2013nza}%
  \BibitemOpen
  \bibfield  {author} {\bibinfo {author} {\bibfnamefont {K.}~\bibnamefont
  {Hebeler}}, \bibinfo {author} {\bibfnamefont {J.~M.}\ \bibnamefont
  {Lattimer}}, \bibinfo {author} {\bibfnamefont {C.~J.}\ \bibnamefont
  {Pethick}}, \ and\ \bibinfo {author} {\bibfnamefont {A.}~\bibnamefont
  {Schwenk}},\ }\href {\doibase 10.1088/0004-637X/773/1/11} {\bibfield
  {journal} {\bibinfo  {journal} {Astrophys. J.}\ }\textbf {\bibinfo {volume}
  {773}},\ \bibinfo {pages} {11} (\bibinfo {year} {2013})},\ \Eprint
  {http://arxiv.org/abs/1303.4662} {arXiv:1303.4662 [astro-ph.SR]} \BibitemShut
  {NoStop}%
\bibitem [{\citenamefont {Tews}\ \emph
  {et~al.}(2018{\natexlab{b}})\citenamefont {Tews}, \citenamefont {Carlson},
  \citenamefont {Gandolfi},\ and\ \citenamefont {Reddy}}]{Tews:2018kmu}%
  \BibitemOpen
  \bibfield  {author} {\bibinfo {author} {\bibfnamefont {I.}~\bibnamefont
  {Tews}}, \bibinfo {author} {\bibfnamefont {J.}~\bibnamefont {Carlson}},
  \bibinfo {author} {\bibfnamefont {S.}~\bibnamefont {Gandolfi}}, \ and\
  \bibinfo {author} {\bibfnamefont {S.}~\bibnamefont {Reddy}},\ }\href
  {\doibase 10.3847/1538-4357/aac267} {\bibfield  {journal} {\bibinfo
  {journal} {Astrophys. J.}\ }\textbf {\bibinfo {volume} {860}},\ \bibinfo
  {pages} {149} (\bibinfo {year} {2018}{\natexlab{b}})},\ \Eprint
  {http://arxiv.org/abs/1801.01923} {arXiv:1801.01923 [nucl-th]} \BibitemShut
  {NoStop}%
\bibitem [{\citenamefont {Drischler}\ \emph {et~al.}(2020)\citenamefont
  {Drischler}, \citenamefont {Furnstahl}, \citenamefont {Melendez},\ and\
  \citenamefont {Phillips}}]{Drischler:2020hwi}%
  \BibitemOpen
  \bibfield  {author} {\bibinfo {author} {\bibfnamefont {C.}~\bibnamefont
  {Drischler}}, \bibinfo {author} {\bibfnamefont {R.~J.}\ \bibnamefont
  {Furnstahl}}, \bibinfo {author} {\bibfnamefont {J.~A.}\ \bibnamefont
  {Melendez}}, \ and\ \bibinfo {author} {\bibfnamefont {D.~R.}\ \bibnamefont
  {Phillips}},\ }\href {\doibase 10.1103/PhysRevLett.125.202702} {\bibfield
  {journal} {\bibinfo  {journal} {Phys. Rev. Lett.}\ }\textbf {\bibinfo
  {volume} {125}},\ \bibinfo {pages} {202702} (\bibinfo {year} {2020})},\
  \Eprint {http://arxiv.org/abs/2004.07232} {arXiv:2004.07232 [nucl-th]}
  \BibitemShut {NoStop}%
\bibitem [{\citenamefont {Keller}\ \emph {et~al.}(2023)\citenamefont {Keller},
  \citenamefont {Hebeler},\ and\ \citenamefont {Schwenk}}]{Keller:2022crb}%
  \BibitemOpen
  \bibfield  {author} {\bibinfo {author} {\bibfnamefont {J.}~\bibnamefont
  {Keller}}, \bibinfo {author} {\bibfnamefont {K.}~\bibnamefont {Hebeler}}, \
  and\ \bibinfo {author} {\bibfnamefont {A.}~\bibnamefont {Schwenk}},\ }\href
  {\doibase 10.1103/PhysRevLett.130.072701} {\bibfield  {journal} {\bibinfo
  {journal} {Phys. Rev. Lett.}\ }\textbf {\bibinfo {volume} {130}},\ \bibinfo
  {pages} {072701} (\bibinfo {year} {2023})},\ \Eprint
  {http://arxiv.org/abs/2204.14016} {arXiv:2204.14016 [nucl-th]} \BibitemShut
  {NoStop}%
\bibitem [{\citenamefont {Gorda}\ \emph {et~al.}(2018)\citenamefont {Gorda},
  \citenamefont {Kurkela}, \citenamefont {Romatschke}, \citenamefont
  {S\"appi},\ and\ \citenamefont {Vuorinen}}]{Gorda:2018gpy}%
  \BibitemOpen
  \bibfield  {author} {\bibinfo {author} {\bibfnamefont {T.}~\bibnamefont
  {Gorda}}, \bibinfo {author} {\bibfnamefont {A.}~\bibnamefont {Kurkela}},
  \bibinfo {author} {\bibfnamefont {P.}~\bibnamefont {Romatschke}}, \bibinfo
  {author} {\bibfnamefont {M.}~\bibnamefont {S\"appi}}, \ and\ \bibinfo
  {author} {\bibfnamefont {A.}~\bibnamefont {Vuorinen}},\ }\href {\doibase
  10.1103/PhysRevLett.121.202701} {\bibfield  {journal} {\bibinfo  {journal}
  {Phys. Rev. Lett.}\ }\textbf {\bibinfo {volume} {121}},\ \bibinfo {pages}
  {202701} (\bibinfo {year} {2018})},\ \Eprint
  {http://arxiv.org/abs/1807.04120} {arXiv:1807.04120 [hep-ph]} \BibitemShut
  {NoStop}%
\bibitem [{\citenamefont {Gorda}\ \emph {et~al.}(2021)\citenamefont {Gorda},
  \citenamefont {Kurkela}, \citenamefont {Paatelainen}, \citenamefont
  {S\"appi},\ and\ \citenamefont {Vuorinen}}]{Gorda:2021znl}%
  \BibitemOpen
  \bibfield  {author} {\bibinfo {author} {\bibfnamefont {T.}~\bibnamefont
  {Gorda}}, \bibinfo {author} {\bibfnamefont {A.}~\bibnamefont {Kurkela}},
  \bibinfo {author} {\bibfnamefont {R.}~\bibnamefont {Paatelainen}}, \bibinfo
  {author} {\bibfnamefont {S.}~\bibnamefont {S\"appi}}, \ and\ \bibinfo
  {author} {\bibfnamefont {A.}~\bibnamefont {Vuorinen}},\ }\href {\doibase
  10.1103/PhysRevLett.127.162003} {\bibfield  {journal} {\bibinfo  {journal}
  {Phys. Rev. Lett.}\ }\textbf {\bibinfo {volume} {127}},\ \bibinfo {pages}
  {162003} (\bibinfo {year} {2021})},\ \Eprint
  {http://arxiv.org/abs/2103.05658} {arXiv:2103.05658 [hep-ph]} \BibitemShut
  {NoStop}%
\bibitem [{\citenamefont {Gorda}\ \emph
  {et~al.}(2023{\natexlab{a}})\citenamefont {Gorda}, \citenamefont
  {Paatelainen}, \citenamefont {S\"appi},\ and\ \citenamefont
  {Sepp\"anen}}]{Gorda:2023mkk}%
  \BibitemOpen
  \bibfield  {author} {\bibinfo {author} {\bibfnamefont {T.}~\bibnamefont
  {Gorda}}, \bibinfo {author} {\bibfnamefont {R.}~\bibnamefont {Paatelainen}},
  \bibinfo {author} {\bibfnamefont {S.}~\bibnamefont {S\"appi}}, \ and\
  \bibinfo {author} {\bibfnamefont {K.}~\bibnamefont {Sepp\"anen}},\
  }\href@noop {} {\  (\bibinfo {year} {2023}{\natexlab{a}})},\ \Eprint
  {http://arxiv.org/abs/2307.08734} {arXiv:2307.08734 [hep-ph]} \BibitemShut
  {NoStop}%
\bibitem [{\citenamefont {Bedaque}\ and\ \citenamefont
  {Steiner}(2015)}]{Bedaque:2014sqa}%
  \BibitemOpen
  \bibfield  {author} {\bibinfo {author} {\bibfnamefont {P.}~\bibnamefont
  {Bedaque}}\ and\ \bibinfo {author} {\bibfnamefont {A.~W.}\ \bibnamefont
  {Steiner}},\ }\href {\doibase 10.1103/PhysRevLett.114.031103} {\bibfield
  {journal} {\bibinfo  {journal} {Phys. Rev. Lett.}\ }\textbf {\bibinfo
  {volume} {114}},\ \bibinfo {pages} {031103} (\bibinfo {year} {2015})},\
  \Eprint {http://arxiv.org/abs/1408.5116} {arXiv:1408.5116 [nucl-th]}
  \BibitemShut {NoStop}%
\bibitem [{\citenamefont {McLerran}\ and\ \citenamefont
  {Reddy}(2019)}]{McLerran:2018hbz}%
  \BibitemOpen
  \bibfield  {author} {\bibinfo {author} {\bibfnamefont {L.}~\bibnamefont
  {McLerran}}\ and\ \bibinfo {author} {\bibfnamefont {S.}~\bibnamefont
  {Reddy}},\ }\href {\doibase 10.1103/PhysRevLett.122.122701} {\bibfield
  {journal} {\bibinfo  {journal} {Phys. Rev. Lett.}\ }\textbf {\bibinfo
  {volume} {122}},\ \bibinfo {pages} {122701} (\bibinfo {year} {2019})},\
  \Eprint {http://arxiv.org/abs/1811.12503} {arXiv:1811.12503 [nucl-th]}
  \BibitemShut {NoStop}%
\bibitem [{\citenamefont {Jeong}\ \emph {et~al.}(2020)\citenamefont {Jeong},
  \citenamefont {McLerran},\ and\ \citenamefont {Sen}}]{Jeong:2019lhv}%
  \BibitemOpen
  \bibfield  {author} {\bibinfo {author} {\bibfnamefont {K.~S.}\ \bibnamefont
  {Jeong}}, \bibinfo {author} {\bibfnamefont {L.}~\bibnamefont {McLerran}}, \
  and\ \bibinfo {author} {\bibfnamefont {S.}~\bibnamefont {Sen}},\ }\href
  {\doibase 10.1103/PhysRevC.101.035201} {\bibfield  {journal} {\bibinfo
  {journal} {Phys. Rev. C}\ }\textbf {\bibinfo {volume} {101}},\ \bibinfo
  {pages} {035201} (\bibinfo {year} {2020})},\ \Eprint
  {http://arxiv.org/abs/1908.04799} {arXiv:1908.04799 [nucl-th]} \BibitemShut
  {NoStop}%
\bibitem [{\citenamefont {Sen}\ and\ \citenamefont
  {Sivertsen}(2021)}]{Sen:2020qcd}%
  \BibitemOpen
  \bibfield  {author} {\bibinfo {author} {\bibfnamefont {S.}~\bibnamefont
  {Sen}}\ and\ \bibinfo {author} {\bibfnamefont {L.}~\bibnamefont
  {Sivertsen}},\ }\href {\doibase 10.3847/1538-4357/abff4c} {\bibfield
  {journal} {\bibinfo  {journal} {Astrophys. J.}\ }\textbf {\bibinfo {volume}
  {915}},\ \bibinfo {pages} {109} (\bibinfo {year} {2021})},\ \Eprint
  {http://arxiv.org/abs/2011.04681} {arXiv:2011.04681 [astro-ph.HE]}
  \BibitemShut {NoStop}%
\bibitem [{\citenamefont {Abbott}\ \emph {et~al.}(2017)\citenamefont {Abbott}
  \emph {et~al.}}]{TheLIGOScientific:2017qsa}%
  \BibitemOpen
  \bibfield  {author} {\bibinfo {author} {\bibfnamefont {B.~P.}\ \bibnamefont
  {Abbott}} \emph {et~al.} (\bibinfo {collaboration} {Virgo, LIGO
  Scientific}),\ }\href {\doibase 10.1103/PhysRevLett.119.161101} {\bibfield
  {journal} {\bibinfo  {journal} {Phys. Rev. Lett.}\ }\textbf {\bibinfo
  {volume} {119}},\ \bibinfo {pages} {161101} (\bibinfo {year} {2017})},\
  \Eprint {http://arxiv.org/abs/1710.05832} {arXiv:1710.05832 [gr-qc]}
  \BibitemShut {NoStop}%
\bibitem [{\citenamefont {Abbott}\ \emph {et~al.}(2020)\citenamefont {Abbott}
  \emph {et~al.}}]{Abbott:2020uma}%
  \BibitemOpen
  \bibfield  {author} {\bibinfo {author} {\bibfnamefont {B.}~\bibnamefont
  {Abbott}} \emph {et~al.} (\bibinfo {collaboration} {LIGO Scientific,
  Virgo}),\ }\href {\doibase 10.3847/2041-8213/ab75f5} {\bibfield  {journal}
  {\bibinfo  {journal} {Astrophys. J. Lett.}\ }\textbf {\bibinfo {volume}
  {892}},\ \bibinfo {pages} {L3} (\bibinfo {year} {2020})},\ \Eprint
  {http://arxiv.org/abs/2001.01761} {arXiv:2001.01761 [astro-ph.HE]}
  \BibitemShut {NoStop}%
\bibitem [{\citenamefont {Aasi}\ \emph {et~al.}(2015)\citenamefont {Aasi} \emph
  {et~al.}}]{LIGOScientific:2014pky}%
  \BibitemOpen
  \bibfield  {author} {\bibinfo {author} {\bibfnamefont {J.}~\bibnamefont
  {Aasi}} \emph {et~al.} (\bibinfo {collaboration} {LIGO Scientific}),\ }\href
  {\doibase 10.1088/0264-9381/32/7/074001} {\bibfield  {journal} {\bibinfo
  {journal} {Class. Quant. Grav.}\ }\textbf {\bibinfo {volume} {32}},\ \bibinfo
  {pages} {074001} (\bibinfo {year} {2015})},\ \Eprint
  {http://arxiv.org/abs/1411.4547} {arXiv:1411.4547 [gr-qc]} \BibitemShut
  {NoStop}%
\bibitem [{\citenamefont {Acernese}\ \emph {et~al.}(2015)\citenamefont
  {Acernese} \emph {et~al.}}]{VIRGO:2014yos}%
  \BibitemOpen
  \bibfield  {author} {\bibinfo {author} {\bibfnamefont {F.}~\bibnamefont
  {Acernese}} \emph {et~al.} (\bibinfo {collaboration} {VIRGO}),\ }\href
  {\doibase 10.1088/0264-9381/32/2/024001} {\bibfield  {journal} {\bibinfo
  {journal} {Class. Quant. Grav.}\ }\textbf {\bibinfo {volume} {32}},\ \bibinfo
  {pages} {024001} (\bibinfo {year} {2015})},\ \Eprint
  {http://arxiv.org/abs/1408.3978} {arXiv:1408.3978 [gr-qc]} \BibitemShut
  {NoStop}%
\bibitem [{GBM(2017)}]{GBM:2017lvd}%
  \BibitemOpen
  \href {\doibase 10.3847/2041-8213/aa91c9} {\bibfield  {journal} {\bibinfo
  {journal} {Astrophys. J.}\ }\textbf {\bibinfo {volume} {848}},\ \bibinfo
  {pages} {L12} (\bibinfo {year} {2017})},\ \Eprint
  {http://arxiv.org/abs/1710.05833} {arXiv:1710.05833 [astro-ph.HE]}
  \BibitemShut {NoStop}%
\bibitem [{\citenamefont {Miller}\ \emph {et~al.}(2019)\citenamefont {Miller}
  \emph {et~al.}}]{Miller:2019cac}%
  \BibitemOpen
  \bibfield  {author} {\bibinfo {author} {\bibfnamefont {M.~C.}\ \bibnamefont
  {Miller}} \emph {et~al.},\ }\href {\doibase 10.3847/2041-8213/ab50c5}
  {\bibfield  {journal} {\bibinfo  {journal} {Astrophys. J. Lett.}\ }\textbf
  {\bibinfo {volume} {887}},\ \bibinfo {pages} {L24} (\bibinfo {year}
  {2019})},\ \Eprint {http://arxiv.org/abs/1912.05705} {arXiv:1912.05705
  [astro-ph.HE]} \BibitemShut {NoStop}%
\bibitem [{\citenamefont {Riley}\ \emph {et~al.}(2019)\citenamefont {Riley}
  \emph {et~al.}}]{Riley:2019yda}%
  \BibitemOpen
  \bibfield  {author} {\bibinfo {author} {\bibfnamefont {T.~E.}\ \bibnamefont
  {Riley}} \emph {et~al.},\ }\href {\doibase 10.3847/2041-8213/ab481c}
  {\bibfield  {journal} {\bibinfo  {journal} {Astrophys. J. Lett.}\ }\textbf
  {\bibinfo {volume} {887}},\ \bibinfo {pages} {L21} (\bibinfo {year}
  {2019})},\ \Eprint {http://arxiv.org/abs/1912.05702} {arXiv:1912.05702
  [astro-ph.HE]} \BibitemShut {NoStop}%
\bibitem [{\citenamefont {Miller}\ \emph {et~al.}(2021)\citenamefont {Miller}
  \emph {et~al.}}]{Miller:2021qha}%
  \BibitemOpen
  \bibfield  {author} {\bibinfo {author} {\bibfnamefont {M.~C.}\ \bibnamefont
  {Miller}} \emph {et~al.},\ }\href {\doibase 10.3847/2041-8213/ac089b}
  {\bibfield  {journal} {\bibinfo  {journal} {Astrophys. J. Lett.}\ }\textbf
  {\bibinfo {volume} {918}},\ \bibinfo {pages} {L28} (\bibinfo {year}
  {2021})},\ \Eprint {http://arxiv.org/abs/2105.06979} {arXiv:2105.06979
  [astro-ph.HE]} \BibitemShut {NoStop}%
\bibitem [{\citenamefont {Riley}\ \emph {et~al.}(2021)\citenamefont {Riley}
  \emph {et~al.}}]{Riley:2021pdl}%
  \BibitemOpen
  \bibfield  {author} {\bibinfo {author} {\bibfnamefont {T.~E.}\ \bibnamefont
  {Riley}} \emph {et~al.},\ }\href {\doibase 10.3847/2041-8213/ac0a81}
  {\bibfield  {journal} {\bibinfo  {journal} {Astrophys. J. Lett.}\ }\textbf
  {\bibinfo {volume} {918}},\ \bibinfo {pages} {L27} (\bibinfo {year}
  {2021})},\ \Eprint {http://arxiv.org/abs/2105.06980} {arXiv:2105.06980
  [astro-ph.HE]} \BibitemShut {NoStop}%
\bibitem [{\citenamefont {Arzoumanian}\ \emph {et~al.}(2018)\citenamefont
  {Arzoumanian} \emph {et~al.}}]{Arzoumanian:2017puf}%
  \BibitemOpen
  \bibfield  {author} {\bibinfo {author} {\bibfnamefont {Z.}~\bibnamefont
  {Arzoumanian}} \emph {et~al.} (\bibinfo {collaboration} {NANOGrav}),\ }\href
  {\doibase 10.3847/1538-4365/aab5b0} {\bibfield  {journal} {\bibinfo
  {journal} {Astrophys. J. Suppl.}\ }\textbf {\bibinfo {volume} {235}},\
  \bibinfo {pages} {37} (\bibinfo {year} {2018})},\ \Eprint
  {http://arxiv.org/abs/1801.01837} {arXiv:1801.01837 [astro-ph.HE]}
  \BibitemShut {NoStop}%
\bibitem [{\citenamefont {Margueron}\ \emph {et~al.}(2021)\citenamefont
  {Margueron}, \citenamefont {Hansen}, \citenamefont {Proust},\ and\
  \citenamefont {Chanfray}}]{PhysRevC.104.055803}%
  \BibitemOpen
  \bibfield  {author} {\bibinfo {author} {\bibfnamefont {J.}~\bibnamefont
  {Margueron}}, \bibinfo {author} {\bibfnamefont {H.}~\bibnamefont {Hansen}},
  \bibinfo {author} {\bibfnamefont {P.}~\bibnamefont {Proust}}, \ and\ \bibinfo
  {author} {\bibfnamefont {G.}~\bibnamefont {Chanfray}},\ }\href {\doibase
  10.1103/PhysRevC.104.055803} {\bibfield  {journal} {\bibinfo  {journal}
  {Phys. Rev. C}\ }\textbf {\bibinfo {volume} {104}},\ \bibinfo {pages}
  {055803} (\bibinfo {year} {2021})}\BibitemShut {NoStop}%
\bibitem [{\citenamefont {Zhao}\ and\ \citenamefont
  {Lattimer}(2020)}]{PhysRevD.102.023021}%
  \BibitemOpen
  \bibfield  {author} {\bibinfo {author} {\bibfnamefont {T.}~\bibnamefont
  {Zhao}}\ and\ \bibinfo {author} {\bibfnamefont {J.~M.}\ \bibnamefont
  {Lattimer}},\ }\href {\doibase 10.1103/PhysRevD.102.023021} {\bibfield
  {journal} {\bibinfo  {journal} {Phys. Rev. D}\ }\textbf {\bibinfo {volume}
  {102}},\ \bibinfo {pages} {023021} (\bibinfo {year} {2020})}\BibitemShut
  {NoStop}%
\bibitem [{\citenamefont {Duarte}\ \emph
  {et~al.}(2020{\natexlab{a}})\citenamefont {Duarte}, \citenamefont
  {Hernandez-Ortiz},\ and\ \citenamefont {Jeong}}]{Duarte:2020xsp}%
  \BibitemOpen
  \bibfield  {author} {\bibinfo {author} {\bibfnamefont {D.~C.}\ \bibnamefont
  {Duarte}}, \bibinfo {author} {\bibfnamefont {S.}~\bibnamefont
  {Hernandez-Ortiz}}, \ and\ \bibinfo {author} {\bibfnamefont {K.~S.}\
  \bibnamefont {Jeong}},\ }\href {\doibase 10.1103/PhysRevC.102.025203}
  {\bibfield  {journal} {\bibinfo  {journal} {Phys. Rev. C}\ }\textbf {\bibinfo
  {volume} {102}},\ \bibinfo {pages} {025203} (\bibinfo {year}
  {2020}{\natexlab{a}})},\ \Eprint {http://arxiv.org/abs/2003.02362}
  {arXiv:2003.02362 [nucl-th]} \BibitemShut {NoStop}%
\bibitem [{\citenamefont {Duarte}\ \emph
  {et~al.}(2020{\natexlab{b}})\citenamefont {Duarte}, \citenamefont
  {Hernandez-Ortiz},\ and\ \citenamefont {Jeong}}]{Duarte_2020}%
  \BibitemOpen
  \bibfield  {author} {\bibinfo {author} {\bibfnamefont {D.~C.}\ \bibnamefont
  {Duarte}}, \bibinfo {author} {\bibfnamefont {S.}~\bibnamefont
  {Hernandez-Ortiz}}, \ and\ \bibinfo {author} {\bibfnamefont {K.~S.}\
  \bibnamefont {Jeong}},\ }\href {\doibase 10.1103/physrevc.102.065202}
  {\bibfield  {journal} {\bibinfo  {journal} {Physical Review C}\ }\textbf
  {\bibinfo {volume} {102}} (\bibinfo {year} {2020}{\natexlab{b}}),\
  10.1103/physrevc.102.065202}\BibitemShut {NoStop}%
\bibitem [{\citenamefont {Kumar}\ \emph {et~al.}(2023)\citenamefont {Kumar},
  \citenamefont {Dey}, \citenamefont {Haque}, \citenamefont {Mallick},\ and\
  \citenamefont {Patra}}]{kumar2023quarkyonic}%
  \BibitemOpen
  \bibfield  {author} {\bibinfo {author} {\bibfnamefont {A.}~\bibnamefont
  {Kumar}}, \bibinfo {author} {\bibfnamefont {D.}~\bibnamefont {Dey}}, \bibinfo
  {author} {\bibfnamefont {S.}~\bibnamefont {Haque}}, \bibinfo {author}
  {\bibfnamefont {R.}~\bibnamefont {Mallick}}, \ and\ \bibinfo {author}
  {\bibfnamefont {S.~K.}\ \bibnamefont {Patra}},\ }\href@noop {} {\enquote
  {\bibinfo {title} {Quarkyonic model for neutron star matter: A relativistic
  mean-field approach},}\ } (\bibinfo {year} {2023}),\ \Eprint
  {http://arxiv.org/abs/2304.08223} {arXiv:2304.08223 [nucl-th]} \BibitemShut
  {NoStop}%
\bibitem [{\citenamefont {Xia}\ \emph {et~al.}(2023)\citenamefont {Xia},
  \citenamefont {Jin},\ and\ \citenamefont {Sun}}]{xia2023quarkyonic}%
  \BibitemOpen
  \bibfield  {author} {\bibinfo {author} {\bibfnamefont {C.-J.}\ \bibnamefont
  {Xia}}, \bibinfo {author} {\bibfnamefont {H.-M.}\ \bibnamefont {Jin}}, \ and\
  \bibinfo {author} {\bibfnamefont {T.-T.}\ \bibnamefont {Sun}},\ }\href@noop
  {} {\enquote {\bibinfo {title} {Quarkyonic matter and quarkyonic stars in an
  extended rmf model},}\ } (\bibinfo {year} {2023}),\ \Eprint
  {http://arxiv.org/abs/2307.03032} {arXiv:2307.03032 [hep-ph]} \BibitemShut
  {NoStop}%
\bibitem [{\citenamefont {Dietrich}\ \emph {et~al.}(2020)\citenamefont
  {Dietrich}, \citenamefont {Coughlin}, \citenamefont {Pang}, \citenamefont
  {Bulla}, \citenamefont {Heinzel}, \citenamefont {Issa}, \citenamefont
  {Tews},\ and\ \citenamefont {Antier}}]{Dietrich:2020lps}%
  \BibitemOpen
  \bibfield  {author} {\bibinfo {author} {\bibfnamefont {T.}~\bibnamefont
  {Dietrich}}, \bibinfo {author} {\bibfnamefont {M.~W.}\ \bibnamefont
  {Coughlin}}, \bibinfo {author} {\bibfnamefont {P.~T.~H.}\ \bibnamefont
  {Pang}}, \bibinfo {author} {\bibfnamefont {M.}~\bibnamefont {Bulla}},
  \bibinfo {author} {\bibfnamefont {J.}~\bibnamefont {Heinzel}}, \bibinfo
  {author} {\bibfnamefont {L.}~\bibnamefont {Issa}}, \bibinfo {author}
  {\bibfnamefont {I.}~\bibnamefont {Tews}}, \ and\ \bibinfo {author}
  {\bibfnamefont {S.}~\bibnamefont {Antier}},\ }\href {\doibase
  10.1126/science.abb4317} {\bibfield  {journal} {\bibinfo  {journal}
  {Science}\ }\textbf {\bibinfo {volume} {370}},\ \bibinfo {pages} {1450}
  (\bibinfo {year} {2020})},\ \Eprint {http://arxiv.org/abs/2002.11355}
  {arXiv:2002.11355 [astro-ph.HE]} \BibitemShut {NoStop}%
\bibitem [{\citenamefont {Pang}\ \emph {et~al.}(2022)\citenamefont {Pang} \emph
  {et~al.}}]{Pang:2022rzc}%
  \BibitemOpen
  \bibfield  {author} {\bibinfo {author} {\bibfnamefont {P.~T.~H.}\
  \bibnamefont {Pang}} \emph {et~al.},\ }\href@noop {} {\  (\bibinfo {year}
  {2022})},\ \Eprint {http://arxiv.org/abs/2205.08513} {arXiv:2205.08513
  [astro-ph.HE]} \BibitemShut {NoStop}%
\bibitem [{\citenamefont {Huth}\ \emph {et~al.}(2021)\citenamefont {Huth} \emph
  {et~al.}}]{Huth:2021bsp}%
  \BibitemOpen
  \bibfield  {author} {\bibinfo {author} {\bibfnamefont {S.}~\bibnamefont
  {Huth}} \emph {et~al.},\ }\href@noop {} {\bibfield  {journal} {\bibinfo
  {journal} {arXiv:2107.06229}\ } (\bibinfo {year} {2021})}\BibitemShut
  {NoStop}%
\bibitem [{\citenamefont {Lackey}\ and\ \citenamefont
  {Wade}(2015)}]{Lackey:2014fwa}%
  \BibitemOpen
  \bibfield  {author} {\bibinfo {author} {\bibfnamefont {B.~D.}\ \bibnamefont
  {Lackey}}\ and\ \bibinfo {author} {\bibfnamefont {L.}~\bibnamefont {Wade}},\
  }\href {\doibase 10.1103/PhysRevD.91.043002} {\bibfield  {journal} {\bibinfo
  {journal} {Phys.Rev.}\ }\textbf {\bibinfo {volume} {D91}},\ \bibinfo {pages}
  {043002} (\bibinfo {year} {2015})},\ \Eprint {http://arxiv.org/abs/1410.8866}
  {arXiv:1410.8866 [gr-qc]} \BibitemShut {NoStop}%
\bibitem [{\citenamefont {Essick}\ \emph {et~al.}(2020)\citenamefont {Essick},
  \citenamefont {Tews}, \citenamefont {Landry}, \citenamefont {Reddy},\ and\
  \citenamefont {Holz}}]{Essick:2020flb}%
  \BibitemOpen
  \bibfield  {author} {\bibinfo {author} {\bibfnamefont {R.}~\bibnamefont
  {Essick}}, \bibinfo {author} {\bibfnamefont {I.}~\bibnamefont {Tews}},
  \bibinfo {author} {\bibfnamefont {P.}~\bibnamefont {Landry}}, \bibinfo
  {author} {\bibfnamefont {S.}~\bibnamefont {Reddy}}, \ and\ \bibinfo {author}
  {\bibfnamefont {D.~E.}\ \bibnamefont {Holz}},\ }\href {\doibase
  10.1103/PhysRevC.102.055803} {\bibfield  {journal} {\bibinfo  {journal}
  {Phys. Rev. C}\ }\textbf {\bibinfo {volume} {102}},\ \bibinfo {pages}
  {055803} (\bibinfo {year} {2020})},\ \Eprint
  {http://arxiv.org/abs/2004.07744} {arXiv:2004.07744 [astro-ph.HE]}
  \BibitemShut {NoStop}%
\bibitem [{\citenamefont {Wysocki}\ \emph {et~al.}(2020)\citenamefont
  {Wysocki}, \citenamefont {O'Shaughnessy}, \citenamefont {Wade},\ and\
  \citenamefont {Lange}}]{Wysocki:2020myz}%
  \BibitemOpen
  \bibfield  {author} {\bibinfo {author} {\bibfnamefont {D.}~\bibnamefont
  {Wysocki}}, \bibinfo {author} {\bibfnamefont {R.}~\bibnamefont
  {O'Shaughnessy}}, \bibinfo {author} {\bibfnamefont {L.}~\bibnamefont {Wade}},
  \ and\ \bibinfo {author} {\bibfnamefont {J.}~\bibnamefont {Lange}},\
  }\href@noop {} {\bibfield  {journal} {\bibinfo  {journal} {arXiv:2001.01747}\
  } (\bibinfo {year} {2020})},\ \Eprint {http://arxiv.org/abs/2001.01747}
  {arXiv:2001.01747 [gr-qc]} \BibitemShut {NoStop}%
\bibitem [{\citenamefont {Pang}\ \emph {et~al.}(2020)\citenamefont {Pang},
  \citenamefont {Dietrich}, \citenamefont {Tews},\ and\ \citenamefont {Van
  Den~Broeck}}]{Pang:2020ilf}%
  \BibitemOpen
  \bibfield  {author} {\bibinfo {author} {\bibfnamefont {P.~T.~H.}\
  \bibnamefont {Pang}}, \bibinfo {author} {\bibfnamefont {T.}~\bibnamefont
  {Dietrich}}, \bibinfo {author} {\bibfnamefont {I.}~\bibnamefont {Tews}}, \
  and\ \bibinfo {author} {\bibfnamefont {C.}~\bibnamefont {Van Den~Broeck}},\
  }\href {\doibase 10.1103/PhysRevResearch.2.033514} {\bibfield  {journal}
  {\bibinfo  {journal} {Phys. Rev. Res.}\ }\textbf {\bibinfo {volume} {2}},\
  \bibinfo {pages} {033514} (\bibinfo {year} {2020})},\ \Eprint
  {http://arxiv.org/abs/2006.14936} {arXiv:2006.14936 [astro-ph.HE]}
  \BibitemShut {NoStop}%
\bibitem [{\citenamefont {Feroz}\ \emph {et~al.}(2009)\citenamefont {Feroz},
  \citenamefont {Hobson},\ and\ \citenamefont {Bridges}}]{Feroz_2009}%
  \BibitemOpen
  \bibfield  {author} {\bibinfo {author} {\bibfnamefont {F.}~\bibnamefont
  {Feroz}}, \bibinfo {author} {\bibfnamefont {M.~P.}\ \bibnamefont {Hobson}}, \
  and\ \bibinfo {author} {\bibfnamefont {M.}~\bibnamefont {Bridges}},\ }\href
  {\doibase 10.1111/j.1365-2966.2009.14548.x} {\bibfield  {journal} {\bibinfo
  {journal} {Monthly Notices of the Royal Astronomical Society}\ }\textbf
  {\bibinfo {volume} {398}},\ \bibinfo {pages} {1601–1614} (\bibinfo {year}
  {2009})}\BibitemShut {NoStop}%
\bibitem [{\citenamefont {{Buchner, J.}}\ \emph {et~al.}(2014)\citenamefont
  {{Buchner, J.}}, \citenamefont {{Georgakakis, A.}}, \citenamefont {{Nandra,
  K.}}, \citenamefont {{Hsu, L.}}, \citenamefont {{Rangel, C.}}, \citenamefont
  {{Brightman, M.}}, \citenamefont {{Merloni, A.}}, \citenamefont {{Salvato,
  M.}}, \citenamefont {{Donley, J.}},\ and\ \citenamefont {{Kocevski,
  D.}}}]{pymultinest1}%
  \BibitemOpen
  \bibfield  {author} {\bibinfo {author} {\bibnamefont {{Buchner, J.}}},
  \bibinfo {author} {\bibnamefont {{Georgakakis, A.}}}, \bibinfo {author}
  {\bibnamefont {{Nandra, K.}}}, \bibinfo {author} {\bibnamefont {{Hsu, L.}}},
  \bibinfo {author} {\bibnamefont {{Rangel, C.}}}, \bibinfo {author}
  {\bibnamefont {{Brightman, M.}}}, \bibinfo {author} {\bibnamefont {{Merloni,
  A.}}}, \bibinfo {author} {\bibnamefont {{Salvato, M.}}}, \bibinfo {author}
  {\bibnamefont {{Donley, J.}}}, \ and\ \bibinfo {author} {\bibnamefont
  {{Kocevski, D.}}},\ }\href {\doibase 10.1051/0004-6361/201322971} {\bibfield
  {journal} {\bibinfo  {journal} {A\&A}\ }\textbf {\bibinfo {volume} {564}},\
  \bibinfo {pages} {A125} (\bibinfo {year} {2014})}\BibitemShut {NoStop}%
\bibitem [{\citenamefont {Buchner}()}]{pymultinest2}%
  \BibitemOpen
  \bibfield  {author} {\bibinfo {author} {\bibfnamefont {J.}~\bibnamefont
  {Buchner}},\ }\href
  {https://johannesbuchner.github.io/PyMultiNest/pymultinest.html} {\enquote
  {\bibinfo {title} {Pymultinest 2.9 documentation},}\ }\BibitemShut {NoStop}%
\bibitem [{\citenamefont {Salmi}\ \emph {et~al.}(2022)\citenamefont {Salmi}
  \emph {et~al.}}]{Salmi:2022cgy}%
  \BibitemOpen
  \bibfield  {author} {\bibinfo {author} {\bibfnamefont {T.}~\bibnamefont
  {Salmi}} \emph {et~al.},\ }\href {\doibase 10.3847/1538-4357/ac983d}
  {\bibfield  {journal} {\bibinfo  {journal} {Astrophys. J.}\ }\textbf
  {\bibinfo {volume} {941}},\ \bibinfo {pages} {150} (\bibinfo {year}
  {2022})},\ \Eprint {http://arxiv.org/abs/2209.12840} {arXiv:2209.12840
  [astro-ph.HE]} \BibitemShut {NoStop}%
\bibitem [{\citenamefont {Struder}\ \emph {et~al.}(2001)\citenamefont {Struder}
  \emph {et~al.}}]{Struder:2001bh}%
  \BibitemOpen
  \bibfield  {author} {\bibinfo {author} {\bibfnamefont {L.}~\bibnamefont
  {Struder}} \emph {et~al.},\ }\href {\doibase 10.1051/0004-6361:20000066}
  {\bibfield  {journal} {\bibinfo  {journal} {Astron. Astrophys.}\ }\textbf
  {\bibinfo {volume} {365}},\ \bibinfo {pages} {L18} (\bibinfo {year}
  {2001})}\BibitemShut {NoStop}%
\bibitem [{\citenamefont {Turner}\ \emph {et~al.}(2001)\citenamefont {Turner}
  \emph {et~al.}}]{Turner:2000jy}%
  \BibitemOpen
  \bibfield  {author} {\bibinfo {author} {\bibfnamefont {M.~J.~L.}\
  \bibnamefont {Turner}} \emph {et~al.},\ }\href {\doibase
  10.1051/0004-6361:20000087} {\bibfield  {journal} {\bibinfo  {journal}
  {Astron. Astrophys.}\ }\textbf {\bibinfo {volume} {365}},\ \bibinfo {pages}
  {L27} (\bibinfo {year} {2001})},\ \Eprint
  {http://arxiv.org/abs/astro-ph/0011498} {arXiv:astro-ph/0011498} \BibitemShut
  {NoStop}%
\bibitem [{\citenamefont {Fonseca}\ \emph {et~al.}(2021)\citenamefont {Fonseca}
  \emph {et~al.}}]{Fonseca2021}%
  \BibitemOpen
  \bibfield  {author} {\bibinfo {author} {\bibfnamefont {E.}~\bibnamefont
  {Fonseca}} \emph {et~al.},\ }\href {\doibase 10.3847/2041-8213/ac03b8}
  {\bibfield  {journal} {\bibinfo  {journal} {Astrophys. J. Lett.}\ }\textbf
  {\bibinfo {volume} {915}},\ \bibinfo {pages} {L12} (\bibinfo {year}
  {2021})},\ \Eprint {http://arxiv.org/abs/2104.00880} {arXiv:2104.00880
  [astro-ph.HE]} \BibitemShut {NoStop}%
\bibitem [{\citenamefont {Foley}\ \emph {et~al.}(2020)\citenamefont {Foley},
  \citenamefont {Coulter}, \citenamefont {Kilpatrick}, \citenamefont {Piro},
  \citenamefont {Ramirez-Ruiz},\ and\ \citenamefont {Schwab}}]{Foley:2020kus}%
  \BibitemOpen
  \bibfield  {author} {\bibinfo {author} {\bibfnamefont {R.~J.}\ \bibnamefont
  {Foley}}, \bibinfo {author} {\bibfnamefont {D.~A.}\ \bibnamefont {Coulter}},
  \bibinfo {author} {\bibfnamefont {C.~D.}\ \bibnamefont {Kilpatrick}},
  \bibinfo {author} {\bibfnamefont {A.~L.}\ \bibnamefont {Piro}}, \bibinfo
  {author} {\bibfnamefont {E.}~\bibnamefont {Ramirez-Ruiz}}, \ and\ \bibinfo
  {author} {\bibfnamefont {J.}~\bibnamefont {Schwab}},\ }\href {\doibase
  10.1093/mnras/staa725} {\bibfield  {journal} {\bibinfo  {journal} {Mon. Not.
  Roy. Astron. Soc.}\ }\textbf {\bibinfo {volume} {494}},\ \bibinfo {pages}
  {190} (\bibinfo {year} {2020})},\ \Eprint {http://arxiv.org/abs/2002.00956}
  {arXiv:2002.00956 [astro-ph.HE]} \BibitemShut {NoStop}%
\bibitem [{\citenamefont {Han}\ \emph {et~al.}(2020)\citenamefont {Han},
  \citenamefont {Tang}, \citenamefont {Hu}, \citenamefont {Li}, \citenamefont
  {Jiang}, \citenamefont {Jin}, \citenamefont {Fan},\ and\ \citenamefont
  {Wei}}]{Han:2020qmn}%
  \BibitemOpen
  \bibfield  {author} {\bibinfo {author} {\bibfnamefont {M.-Z.}\ \bibnamefont
  {Han}}, \bibinfo {author} {\bibfnamefont {S.-P.}\ \bibnamefont {Tang}},
  \bibinfo {author} {\bibfnamefont {Y.-M.}\ \bibnamefont {Hu}}, \bibinfo
  {author} {\bibfnamefont {Y.-J.}\ \bibnamefont {Li}}, \bibinfo {author}
  {\bibfnamefont {J.-L.}\ \bibnamefont {Jiang}}, \bibinfo {author}
  {\bibfnamefont {Z.-P.}\ \bibnamefont {Jin}}, \bibinfo {author} {\bibfnamefont
  {Y.-Z.}\ \bibnamefont {Fan}}, \ and\ \bibinfo {author} {\bibfnamefont
  {D.-M.}\ \bibnamefont {Wei}},\ }\href {\doibase 10.3847/2041-8213/ab745a}
  {\bibfield  {journal} {\bibinfo  {journal} {Astrophys. J. Lett.}\ }\textbf
  {\bibinfo {volume} {891}},\ \bibinfo {pages} {L5} (\bibinfo {year} {2020})},\
  \Eprint {http://arxiv.org/abs/2001.07882} {arXiv:2001.07882 [astro-ph.HE]}
  \BibitemShut {NoStop}%
\bibitem [{\citenamefont {Kyutoku}\ \emph {et~al.}(2020)\citenamefont
  {Kyutoku}, \citenamefont {Fujibayashi}, \citenamefont {Hayashi},
  \citenamefont {Kawaguchi}, \citenamefont {Kiuchi}, \citenamefont {Shibata},\
  and\ \citenamefont {Tanaka}}]{Kyutoku:2020xka}%
  \BibitemOpen
  \bibfield  {author} {\bibinfo {author} {\bibfnamefont {K.}~\bibnamefont
  {Kyutoku}}, \bibinfo {author} {\bibfnamefont {S.}~\bibnamefont
  {Fujibayashi}}, \bibinfo {author} {\bibfnamefont {K.}~\bibnamefont
  {Hayashi}}, \bibinfo {author} {\bibfnamefont {K.}~\bibnamefont {Kawaguchi}},
  \bibinfo {author} {\bibfnamefont {K.}~\bibnamefont {Kiuchi}}, \bibinfo
  {author} {\bibfnamefont {M.}~\bibnamefont {Shibata}}, \ and\ \bibinfo
  {author} {\bibfnamefont {M.}~\bibnamefont {Tanaka}},\ }\href {\doibase
  10.3847/2041-8213/ab6e70} {\bibfield  {journal} {\bibinfo  {journal}
  {Astrophys. J.}\ }\textbf {\bibinfo {volume} {890}},\ \bibinfo {pages} {L4}
  (\bibinfo {year} {2020})},\ \Eprint {http://arxiv.org/abs/2001.04474}
  {arXiv:2001.04474 [astro-ph.HE]} \BibitemShut {NoStop}%
\bibitem [{\citenamefont {Collaboration}\ and\ \citenamefont {the
  Virgo~Collaboration}(2019)}]{GW170817_PE_samples}%
  \BibitemOpen
  \bibfield  {author} {\bibinfo {author} {\bibfnamefont {T.~L.~S.}\
  \bibnamefont {Collaboration}}\ and\ \bibinfo {author} {\bibnamefont {the
  Virgo~Collaboration}},\ }\href {https://dcc.ligo.org/LIGO-P1800370/public}
  {\enquote {\bibinfo {title} {Parameter estimation sample release for
  gwtc-1},}\ } (\bibinfo {year} {2019})\BibitemShut {NoStop}%
\bibitem [{\citenamefont {Collaboration}\ and\ \citenamefont {the
  Virgo~Collaboration}(2020)}]{GW190425_PE_samples}%
  \BibitemOpen
  \bibfield  {author} {\bibinfo {author} {\bibfnamefont {T.~L.~S.}\
  \bibnamefont {Collaboration}}\ and\ \bibinfo {author} {\bibnamefont {the
  Virgo~Collaboration}},\ }\href {https://dcc.ligo.org/LIGO-P2000026/public}
  {\enquote {\bibinfo {title} {Parameter estimation sample release for
  gw190425},}\ } (\bibinfo {year} {2020})\BibitemShut {NoStop}%
\bibitem [{\citenamefont {Bulla}(2019)}]{Bulla:2019muo}%
  \BibitemOpen
  \bibfield  {author} {\bibinfo {author} {\bibfnamefont {M.}~\bibnamefont
  {Bulla}},\ }\href {\doibase 10.1093/mnras/stz2495} {\bibfield  {journal}
  {\bibinfo  {journal} {Mon. Not. Roy. Astron. Soc.}\ }\textbf {\bibinfo
  {volume} {489}},\ \bibinfo {pages} {5037} (\bibinfo {year} {2019})},\ \Eprint
  {http://arxiv.org/abs/1906.04205} {arXiv:1906.04205 [astro-ph.HE]}
  \BibitemShut {NoStop}%
\bibitem [{\citenamefont {Kr{\"u}ger}\ and\ \citenamefont
  {Foucart}(2020)}]{Kruger:2020gig}%
  \BibitemOpen
  \bibfield  {author} {\bibinfo {author} {\bibfnamefont {C.~J.}\ \bibnamefont
  {Kr{\"u}ger}}\ and\ \bibinfo {author} {\bibfnamefont {F.}~\bibnamefont
  {Foucart}},\ }\href {\doibase 10.1103/PhysRevD.101.103002} {\bibfield
  {journal} {\bibinfo  {journal} {Phys. Rev. D}\ }\textbf {\bibinfo {volume}
  {101}},\ \bibinfo {pages} {103002} (\bibinfo {year} {2020})},\ \Eprint
  {http://arxiv.org/abs/2002.07728} {arXiv:2002.07728 [astro-ph.HE]}
  \BibitemShut {NoStop}%
\bibitem [{\citenamefont {Bauswein}\ \emph {et~al.}(2020)\citenamefont
  {Bauswein}, \citenamefont {Blacker}, \citenamefont {Vijayan}, \citenamefont
  {Stergioulas}, \citenamefont {Chatziioannou}, \citenamefont {Clark},
  \citenamefont {Bastian}, \citenamefont {Blaschke}, \citenamefont {Cierniak},\
  and\ \citenamefont {Fischer}}]{Bauswein:2020aag}%
  \BibitemOpen
  \bibfield  {author} {\bibinfo {author} {\bibfnamefont {A.}~\bibnamefont
  {Bauswein}}, \bibinfo {author} {\bibfnamefont {S.}~\bibnamefont {Blacker}},
  \bibinfo {author} {\bibfnamefont {V.}~\bibnamefont {Vijayan}}, \bibinfo
  {author} {\bibfnamefont {N.}~\bibnamefont {Stergioulas}}, \bibinfo {author}
  {\bibfnamefont {K.}~\bibnamefont {Chatziioannou}}, \bibinfo {author}
  {\bibfnamefont {J.~A.}\ \bibnamefont {Clark}}, \bibinfo {author}
  {\bibfnamefont {N.-U.~F.}\ \bibnamefont {Bastian}}, \bibinfo {author}
  {\bibfnamefont {D.~B.}\ \bibnamefont {Blaschke}}, \bibinfo {author}
  {\bibfnamefont {M.}~\bibnamefont {Cierniak}}, \ and\ \bibinfo {author}
  {\bibfnamefont {T.}~\bibnamefont {Fischer}},\ }\href {\doibase
  10.1103/PhysRevLett.125.141103} {\bibfield  {journal} {\bibinfo  {journal}
  {Phys. Rev. Lett.}\ }\textbf {\bibinfo {volume} {125}},\ \bibinfo {pages}
  {141103} (\bibinfo {year} {2020})},\ \Eprint
  {http://arxiv.org/abs/2004.00846} {arXiv:2004.00846 [astro-ph.HE]}
  \BibitemShut {NoStop}%
\bibitem [{\citenamefont {Kataev}\ and\ \citenamefont
  {Molokoedov}(2015)}]{PhysRevD.92.054008}%
  \BibitemOpen
  \bibfield  {author} {\bibinfo {author} {\bibfnamefont {A.~L.}\ \bibnamefont
  {Kataev}}\ and\ \bibinfo {author} {\bibfnamefont {V.~S.}\ \bibnamefont
  {Molokoedov}},\ }\href {\doibase 10.1103/PhysRevD.92.054008} {\bibfield
  {journal} {\bibinfo  {journal} {Phys. Rev. D}\ }\textbf {\bibinfo {volume}
  {92}},\ \bibinfo {pages} {054008} (\bibinfo {year} {2015})}\BibitemShut
  {NoStop}%
\bibitem [{\citenamefont {Essick}\ \emph
  {et~al.}(2021{\natexlab{a}})\citenamefont {Essick}, \citenamefont {Tews},
  \citenamefont {Landry},\ and\ \citenamefont {Schwenk}}]{Essick:2021kjb}%
  \BibitemOpen
  \bibfield  {author} {\bibinfo {author} {\bibfnamefont {R.}~\bibnamefont
  {Essick}}, \bibinfo {author} {\bibfnamefont {I.}~\bibnamefont {Tews}},
  \bibinfo {author} {\bibfnamefont {P.}~\bibnamefont {Landry}}, \ and\ \bibinfo
  {author} {\bibfnamefont {A.}~\bibnamefont {Schwenk}},\ }\href@noop {} {\
  (\bibinfo {year} {2021}{\natexlab{a}})},\ \Eprint
  {http://arxiv.org/abs/2102.10074} {arXiv:2102.10074 [nucl-th]} \BibitemShut
  {NoStop}%
\bibitem [{\citenamefont {Essick}\ \emph
  {et~al.}(2021{\natexlab{b}})\citenamefont {Essick}, \citenamefont {Landry},
  \citenamefont {Schwenk},\ and\ \citenamefont {Tews}}]{Essick:2021ezp}%
  \BibitemOpen
  \bibfield  {author} {\bibinfo {author} {\bibfnamefont {R.}~\bibnamefont
  {Essick}}, \bibinfo {author} {\bibfnamefont {P.}~\bibnamefont {Landry}},
  \bibinfo {author} {\bibfnamefont {A.}~\bibnamefont {Schwenk}}, \ and\
  \bibinfo {author} {\bibfnamefont {I.}~\bibnamefont {Tews}},\ }\href@noop {}
  {\  (\bibinfo {year} {2021}{\natexlab{b}})},\ \Eprint
  {http://arxiv.org/abs/2107.05528} {arXiv:2107.05528 [nucl-th]} \BibitemShut
  {NoStop}%
\bibitem [{\citenamefont {Vinciguerra}\ \emph {et~al.}(2023)\citenamefont
  {Vinciguerra} \emph {et~al.}}]{Vinciguerra:2023qxq}%
  \BibitemOpen
  \bibfield  {author} {\bibinfo {author} {\bibfnamefont {S.}~\bibnamefont
  {Vinciguerra}} \emph {et~al.},\ }\href@noop {} {\  (\bibinfo {year}
  {2023})},\ \Eprint {http://arxiv.org/abs/2308.09469} {arXiv:2308.09469
  [astro-ph.HE]} \BibitemShut {NoStop}%
\bibitem [{\citenamefont {Legred}\ \emph {et~al.}(2021)\citenamefont {Legred},
  \citenamefont {Chatziioannou}, \citenamefont {Essick}, \citenamefont {Han},\
  and\ \citenamefont {Landry}}]{Legred:2021hdx}%
  \BibitemOpen
  \bibfield  {author} {\bibinfo {author} {\bibfnamefont {I.}~\bibnamefont
  {Legred}}, \bibinfo {author} {\bibfnamefont {K.}~\bibnamefont
  {Chatziioannou}}, \bibinfo {author} {\bibfnamefont {R.}~\bibnamefont
  {Essick}}, \bibinfo {author} {\bibfnamefont {S.}~\bibnamefont {Han}}, \ and\
  \bibinfo {author} {\bibfnamefont {P.}~\bibnamefont {Landry}},\ }\href@noop {}
  {\bibfield  {journal} {\bibinfo  {journal} {arXiv:2106.05313}\ } (\bibinfo
  {year} {2021})}\BibitemShut {NoStop}%
\bibitem [{\citenamefont {Fujimoto}\ \emph {et~al.}(2022)\citenamefont
  {Fujimoto}, \citenamefont {Fukushima}, \citenamefont {McLerran},\ and\
  \citenamefont {Praszalowicz}}]{Fujimoto:2022ohj}%
  \BibitemOpen
  \bibfield  {author} {\bibinfo {author} {\bibfnamefont {Y.}~\bibnamefont
  {Fujimoto}}, \bibinfo {author} {\bibfnamefont {K.}~\bibnamefont {Fukushima}},
  \bibinfo {author} {\bibfnamefont {L.~D.}\ \bibnamefont {McLerran}}, \ and\
  \bibinfo {author} {\bibfnamefont {M.}~\bibnamefont {Praszalowicz}},\ }\href
  {\doibase 10.1103/PhysRevLett.129.252702} {\bibfield  {journal} {\bibinfo
  {journal} {Phys. Rev. Lett.}\ }\textbf {\bibinfo {volume} {129}},\ \bibinfo
  {pages} {252702} (\bibinfo {year} {2022})},\ \Eprint
  {http://arxiv.org/abs/2207.06753} {arXiv:2207.06753 [nucl-th]} \BibitemShut
  {NoStop}%
\bibitem [{\citenamefont {Somasundaram}\ \emph {et~al.}(2023)\citenamefont
  {Somasundaram}, \citenamefont {Tews},\ and\ \citenamefont
  {Margueron}}]{Somasundaram:2022ztm}%
  \BibitemOpen
  \bibfield  {author} {\bibinfo {author} {\bibfnamefont {R.}~\bibnamefont
  {Somasundaram}}, \bibinfo {author} {\bibfnamefont {I.}~\bibnamefont {Tews}},
  \ and\ \bibinfo {author} {\bibfnamefont {J.}~\bibnamefont {Margueron}},\
  }\href {\doibase 10.1103/PhysRevC.107.L052801} {\bibfield  {journal}
  {\bibinfo  {journal} {Phys. Rev. C}\ }\textbf {\bibinfo {volume} {107}},\
  \bibinfo {pages} {L052801} (\bibinfo {year} {2023})},\ \Eprint
  {http://arxiv.org/abs/2204.14039} {arXiv:2204.14039 [nucl-th]} \BibitemShut
  {NoStop}%
\bibitem [{\citenamefont {Komoltsev}\ and\ \citenamefont
  {Kurkela}(2022)}]{Komoltsev:2021jzg}%
  \BibitemOpen
  \bibfield  {author} {\bibinfo {author} {\bibfnamefont {O.}~\bibnamefont
  {Komoltsev}}\ and\ \bibinfo {author} {\bibfnamefont {A.}~\bibnamefont
  {Kurkela}},\ }\href {\doibase 10.1103/PhysRevLett.128.202701} {\bibfield
  {journal} {\bibinfo  {journal} {Phys. Rev. Lett.}\ }\textbf {\bibinfo
  {volume} {128}},\ \bibinfo {pages} {202701} (\bibinfo {year} {2022})},\
  \Eprint {http://arxiv.org/abs/2111.05350} {arXiv:2111.05350 [nucl-th]}
  \BibitemShut {NoStop}%
\bibitem [{\citenamefont {Gorda}\ \emph
  {et~al.}(2023{\natexlab{b}})\citenamefont {Gorda}, \citenamefont
  {Komoltsev},\ and\ \citenamefont {Kurkela}}]{Gorda:2022jvk}%
  \BibitemOpen
  \bibfield  {author} {\bibinfo {author} {\bibfnamefont {T.}~\bibnamefont
  {Gorda}}, \bibinfo {author} {\bibfnamefont {O.}~\bibnamefont {Komoltsev}}, \
  and\ \bibinfo {author} {\bibfnamefont {A.}~\bibnamefont {Kurkela}},\ }\href
  {\doibase 10.3847/1538-4357/acce3a} {\bibfield  {journal} {\bibinfo
  {journal} {Astrophys. J.}\ }\textbf {\bibinfo {volume} {950}},\ \bibinfo
  {pages} {107} (\bibinfo {year} {2023}{\natexlab{b}})},\ \Eprint
  {http://arxiv.org/abs/2204.11877} {arXiv:2204.11877 [nucl-th]} \BibitemShut
  {NoStop}%
\bibitem [{\citenamefont {Le~F\`evre}\ \emph {et~al.}(2016)\citenamefont
  {Le~F\`evre}, \citenamefont {Leifels}, \citenamefont {Reisdorf},
  \citenamefont {Aichelin},\ and\ \citenamefont {Hartnack}}]{LeFevre:2015paj}%
  \BibitemOpen
  \bibfield  {author} {\bibinfo {author} {\bibfnamefont {A.}~\bibnamefont
  {Le~F\`evre}}, \bibinfo {author} {\bibfnamefont {Y.}~\bibnamefont {Leifels}},
  \bibinfo {author} {\bibfnamefont {W.}~\bibnamefont {Reisdorf}}, \bibinfo
  {author} {\bibfnamefont {J.}~\bibnamefont {Aichelin}}, \ and\ \bibinfo
  {author} {\bibfnamefont {C.}~\bibnamefont {Hartnack}},\ }\href {\doibase
  10.1016/j.nuclphysa.2015.09.015} {\bibfield  {journal} {\bibinfo  {journal}
  {Nucl. Phys. A}\ }\textbf {\bibinfo {volume} {945}},\ \bibinfo {pages} {112}
  (\bibinfo {year} {2016})},\ \Eprint {http://arxiv.org/abs/1501.05246}
  {arXiv:1501.05246 [nucl-ex]} \BibitemShut {NoStop}%
\bibitem [{\citenamefont {Russotto}\ \emph {et~al.}(2016)\citenamefont
  {Russotto} \emph {et~al.}}]{Russotto:2016ucm}%
  \BibitemOpen
  \bibfield  {author} {\bibinfo {author} {\bibfnamefont {P.}~\bibnamefont
  {Russotto}} \emph {et~al.},\ }\href {\doibase 10.1103/PhysRevC.94.034608}
  {\bibfield  {journal} {\bibinfo  {journal} {Phys. Rev. C}\ }\textbf {\bibinfo
  {volume} {94}},\ \bibinfo {pages} {034608} (\bibinfo {year} {2016})},\
  \Eprint {http://arxiv.org/abs/1608.04332} {arXiv:1608.04332 [nucl-ex]}
  \BibitemShut {NoStop}%
\bibitem [{\citenamefont {Adhikari}\ \emph {et~al.}(2021)\citenamefont
  {Adhikari}, \citenamefont {Albataineh}, \citenamefont {Androic},
  \citenamefont {Aniol}, \citenamefont {Armstrong}, \citenamefont {Averett},
  \citenamefont {Ayerbe~Gayoso}, \citenamefont {Barcus}, \citenamefont
  {Bellini}, \citenamefont {Beminiwattha} \emph {et~al.}}]{PREXII}%
  \BibitemOpen
  \bibfield  {author} {\bibinfo {author} {\bibfnamefont {D.}~\bibnamefont
  {Adhikari}}, \bibinfo {author} {\bibfnamefont {H.}~\bibnamefont
  {Albataineh}}, \bibinfo {author} {\bibfnamefont {D.}~\bibnamefont {Androic}},
  \bibinfo {author} {\bibfnamefont {K.}~\bibnamefont {Aniol}}, \bibinfo
  {author} {\bibfnamefont {D.~S.}\ \bibnamefont {Armstrong}}, \bibinfo {author}
  {\bibfnamefont {T.}~\bibnamefont {Averett}}, \bibinfo {author} {\bibfnamefont
  {C.}~\bibnamefont {Ayerbe~Gayoso}}, \bibinfo {author} {\bibfnamefont
  {S.}~\bibnamefont {Barcus}}, \bibinfo {author} {\bibfnamefont
  {V.}~\bibnamefont {Bellini}}, \bibinfo {author} {\bibfnamefont {R.~S.}\
  \bibnamefont {Beminiwattha}},  \emph {et~al.} (\bibinfo {collaboration}
  {PREX}),\ }\href {\doibase 10.1103/PhysRevLett.126.172502} {\bibfield
  {journal} {\bibinfo  {journal} {Phys. Rev. Lett.}\ }\textbf {\bibinfo
  {volume} {126}},\ \bibinfo {pages} {172502} (\bibinfo {year} {2021})},\
  \Eprint {http://arxiv.org/abs/2102.10767} {arXiv:2102.10767 [nucl-ex]}
  \BibitemShut {NoStop}%
\bibitem [{\citenamefont {Doroshenko}\ \emph {et~al.}(2022)\citenamefont
  {Doroshenko}, \citenamefont {Suleimanov}, \citenamefont {P{\"u}hlhofer},\
  and\ \citenamefont {Santangelo}}]{Doroshenko2022}%
  \BibitemOpen
  \bibfield  {author} {\bibinfo {author} {\bibfnamefont {V.}~\bibnamefont
  {Doroshenko}}, \bibinfo {author} {\bibfnamefont {V.}~\bibnamefont
  {Suleimanov}}, \bibinfo {author} {\bibfnamefont {G.}~\bibnamefont
  {P{\"u}hlhofer}}, \ and\ \bibinfo {author} {\bibfnamefont {A.}~\bibnamefont
  {Santangelo}},\ }\href {\doibase 10.1038/s41550-022-01800-1} {\bibfield
  {journal} {\bibinfo  {journal} {Nature Astronomy}\ }\textbf {\bibinfo
  {volume} {6}},\ \bibinfo {pages} {1444} (\bibinfo {year} {2022})}\BibitemShut
  {NoStop}%
\bibitem [{\citenamefont {Punturo}\ \emph {et~al.}(2010)\citenamefont {Punturo}
  \emph {et~al.}}]{Punturo:2010zza}%
  \BibitemOpen
  \bibfield  {author} {\bibinfo {author} {\bibfnamefont {M.}~\bibnamefont
  {Punturo}} \emph {et~al.},\ }\href {\doibase 10.1088/0264-9381/27/8/084007}
  {\bibfield  {journal} {\bibinfo  {journal} {Class. Quant. Grav.}\ }\textbf
  {\bibinfo {volume} {27}},\ \bibinfo {pages} {084007} (\bibinfo {year}
  {2010})}\BibitemShut {NoStop}%
\bibitem [{\citenamefont {Hild}\ \emph {et~al.}(2011)\citenamefont {Hild} \emph
  {et~al.}}]{Hild:2010id}%
  \BibitemOpen
  \bibfield  {author} {\bibinfo {author} {\bibfnamefont {S.}~\bibnamefont
  {Hild}} \emph {et~al.},\ }\href {\doibase 10.1088/0264-9381/28/9/094013}
  {\bibfield  {journal} {\bibinfo  {journal} {Class. Quant. Grav.}\ }\textbf
  {\bibinfo {volume} {28}},\ \bibinfo {pages} {094013} (\bibinfo {year}
  {2011})},\ \Eprint {http://arxiv.org/abs/1012.0908} {arXiv:1012.0908 [gr-qc]}
  \BibitemShut {NoStop}%
\bibitem [{\citenamefont {Reitze}\ \emph {et~al.}(2019)\citenamefont {Reitze}
  \emph {et~al.}}]{Reitze:2019iox}%
  \BibitemOpen
  \bibfield  {author} {\bibinfo {author} {\bibfnamefont {D.}~\bibnamefont
  {Reitze}} \emph {et~al.},\ }\href@noop {} {\bibfield  {journal} {\bibinfo
  {journal} {Bull. Am. Astron. Soc.}\ }\textbf {\bibinfo {volume} {51}},\
  \bibinfo {pages} {035} (\bibinfo {year} {2019})},\ \Eprint
  {http://arxiv.org/abs/1907.04833} {arXiv:1907.04833 [astro-ph.IM]}
  \BibitemShut {NoStop}%
\end{thebibliography}%

\end{document}